\documentclass[astrosymb, twocolumn]{aastex631}
\usepackage{enumitem}

\newcommand{\candidate}{\textit{Gaia EDR3 2077240046296834304}}

\shorttitle{AASTeX v6.3.1 Sample article}
\shortauthors{Mart\'inez-Palomera et al.}
\graphicspath{{./}{figures/}}

\begin{document}

\title{Kepler Bonus: Aperture Photometry Light Curves of EXBA Sources}

\correspondingauthor{Jorge Martínez-Palomera}
\email{palomera@baeri.org, jorgemarpa@ug.uchile.cl}

\author[0000-0002-7395-4935]{Jorge Martínez-Palomera}
\affiliation{Bay Area Environmental Research Institute, P.O. Box 25, Moffett Field, CA 94035, USA.}
\affiliation{NASA Ames Research Center, Moffett Field, CA, USA}

\author[0000-0002-3385-8391]{Christina Hedges}
\affiliation{Bay Area Environmental Research Institute, P.O. Box 25, Moffett Field, CA 94035, USA.}
\affiliation{NASA Ames Research Center, Moffett Field, CA, USA}

\author[0000-0001-8812-0565]{Joseph E. Rodriguez} 
\affiliation{Department of Physics and Astronomy, Michigan State University, East Lansing, MI 48824, USA}

\author[0000-0002-3306-3484]{Geert Barentsen}
\affiliation{Bay Area Environmental Research Institute, P.O. Box 25, Moffett Field, CA 94035, USA.}
\affiliation{NASA Ames Research Center, Moffett Field, CA, USA}

\author[0000-0003-4206-5649]{Jessie Dotson}
\affiliation{NASA Ames Research Center, Moffett Field, CA, USA}



\begin{abstract}

NASA's \textit{Kepler} mission observed background regions across its field of view for more than three consecutive years using custom designed super apertures (EXBA masks). Since these apertures were designed to capture a region of the sky rather than single targets, the Kepler Science Data Processing pipeline produced Target Pixel Files, but did not produce light curves for the sources within these background regions. In this work we produce light curves for $9,327$ sources observed in the EXBA masks. These light curves are generated using aperture photometry estimated from the instrument's Pixel Response Function (PRF) profile computed from \textit{Kepler}'s full-frame images. The PRF models enable the creation of apertures that follow the characteristic shapes of the PSF in the image and the computation of flux completeness and contamination metrics. The light curves are available at MAST as a High Level Science Product (\texttt{kbonus-apexba}). Alongside this dataset, we present \texttt{kepler-apertures}, a \texttt{Python} library to compute PRF models and use them to perform aperture photometry on \textit{Kepler}-like data. Using light curves from the EXBA masks we found an exoplanet candidate around \candidate consistent with a large planet companion with a $0.81 R_J$ radius. Additionally, we report a catalog of 69 eclipsing binaries.
We encourage the community to exploit this new dataset to perform in depth time domain analysis, such as eclipsing binaries demographic and others.

\end{abstract}



\section{Introduction} \label{sec:intro}

NASA's \textit{Kepler} mission delivered to the community one of the finest time series datasets ever produced.
\textit{Kepler} observed more than $400,000$ target stars \citep{2010Sci...327..977B} \textit{Kepler} found more than $4,000$ exoplanet candidates \citep{Thompson_2018}, observed numerous supernovae from earliest stages of explosion \citep{2015Natur.521..332O, 2016ApJ...820...23G, 2019ApJ...870...12L}, and more than $2,900$ eclipsing binary systems \citep{2016AJ....151...68K}, to name some examples. 
The \textit{Kepler} mission mainly downlinked image data around selected targets \citep[KIC, ][]{2011AJ....142..112B}. 
These target cutouts are centered on the object of interest and usually cover a few pixels around the source. 
\textit{Kepler} observations are organized into seventeen 93-days periods. 
These observation periods are named quarters.
The Kepler Science Data Processing Pipeline \citep{Jenkins_2010}, produced two science products. 
First, the \textit{Kepler} pipeline produced Target Pixel Files (TPFs), which are cutout images of each observed target. 
A TPF contains all observed cadences for the target in a single quarter as well as information on which pixels fall into the aperture of each target star.
Secondly, the Light Curve Files (LCFs), which are flux time series of the target of interest.
The LCFs contain two types of light curves, one is Simple Aperture Photometry (SAP), where the flux is measured within an aperture selected to minimize contamination from background sources. 
The second type are Presearch Data Conditioned Simple Aperture Photometry (PDCSAP) light curves \citep{2012PASP..124.1000S} that corrects the SAP light curves for systematics of the instrument.

\textit{Kepler}'s mission also downlinked single cadence observations of the entire \textit{Kepler} field each month. 
These Full Frame Images (FFIs) contain more than 1 million sources per image. 
They were observed mainly for calibration and diagnostic purposes \citep{2016ksci.rept....1V}.
Although the FFIs covered the entire focal plane of the instrument, the time resolution is not comparable to the main targets observed with a cadence of 1 or 30 minutes.
\cite{Montet_2017} used \textit{Kepler}'s FFIs to study long-term variability of Sun-like stars due to magnetic cycles.
Additionally, two types of custom aperture targets were observed using the 30 min cadence mode. 
First, aperture masks covering two open clusters, NGC 6819 \citep{2010ApJ...713L.182S, 2013MNRAS.430.3472B, 2013AAS...22125036B} and NGC 6791 \citep{2011ApJ...737L..10S, 2012ApJ...757..190C, 2021AAS...23714006M}. 
The second custom masks are the background super apertures, or EXBA masks. 
The EXBA masks cover relatively dark regions of the sky at a common location on all CCD channels, and were observed continuously between quarters 5 and 17. 
The scientific motivation to collect these images was to obtain an unbiased characterization of the background Eclipsing Binary (EB) rate in the \textit{Kepler} field \citep{2018arXiv181012554B}.

The \textit{Kepler} pipeline relied upon aperture photometry to create the flux time series of the targets. 
There are multiple ways to define the shape and the size of a photometric aperture. 
The simplest approach is defining circular apertures that enclose the source flux. 
A circular aperture would work well for isolated sources and where the instrument's Pixel Response Function (PRF) profile is symmetric (typically found in \textit{Kepler} at the center of the focal plane).
But a circular aperture will lose considerable flux where the PRF shape becomes distorted (typically when moving away from the center of the image)
Aperture photometry can be improved by allowing for elliptical or non-circular shapes, e.g. isophote apertures. 
Regardless of shape simple apertures become sub-optimal in crowded regions, where source contamination is the dominant factor.
\cite{2010ApJ...713L..97B} shows that \textit{Kepler}'s PRF profiles are distorted at the edges of the focal plane, with elongated shapes in multiple directions.
Therefore, the simple aperture methods will perform poorly.
The \textit{Kepler} pipeline performed photometry by computing optimized apertures for every target source and providing metrics that characterize the completeness of the flux, and amount of contamination within the aperture.

The EXBA masks were chosen to have an unbiased representation of the \textit{Kepler} field.
The number of Gaia sources (brighter than $20$th magnitude in the $G$ band) per channel mask fluctuates from less populated regions with a few dozens of objects to more crowded ones with a couple of hundred stars.
In this work, we implement aperture photometry by creating custom apertures that follow the PRF profile of the image.
This PRF models are computed using \textit{Kepler}'s FFIs and evaluated on every EXBA source.
Until now, the EXBA masks have not been systematically analyzed and the \textit{Kepler} pipeline did not produce LCFs for observed sources. 
In this work, we produce light curves for more than $9,300$ sources detected in the EXBA masks. 
We follow the procedures introduced by the Linearized Field Deblending (LFD) photometry method \citep{2021AJ....162..107H} to obtain a detailed, yet fast to compute and evaluate, model of the PRF profile for all the channels in the focal plane of \textit{Kepler}. 
Evaluating the PRF models on the EXBA sources provide a model of the object, which is used to define the photometric aperture and compute flux metrics. 
All the light curves produced in this work are publicly available to the community as FITS Light Curve Files and can be accessed throughout the Mikulski Archive for Space Telescopes (MAST) archive\footnote{Kepler Bonus, APEXBA, {\doi{10.17909/t9-d5wy-e535}}}. 
We present \texttt{kepler-apertures}\footnote{\url{https://github.com/jorgemarpa/kepler-apertures/releases/tag/v0.1.0}} \citep{jorge_martinez_palomera_2021_5062871} a the set of \texttt{Python} tools to compute and use PRF models to perform aperture photometry on \textit{Kepler}-like data.
As an example of the opportunities that this new light-curve dataset provides, we perform a search of exoplanet transiting signals using the BLS periodogram method. 
We identify a candidate around the source \candidate that shows a transiting signal consistent with a large planet or sub-stellar companion. 
Our search also provide a catalog of 69 EBs observed in the EXBA masks.

This article is structured as follows. 
Section \ref{sec:data} details the characteristics of the data used for this work as well as the steps followed to compute the PRF models, aperture photometry, flux metrics, and light curves. 
In Section \ref{sec:results} we present our results: high-level science products consisting of light curve files, and the search for transiting signals which resulted in an exoplanet candidate and a catalog of EBs. 
In Section \ref{sec:discurssion} we discuss the opportunities that this new unexplored dataset provides to the community, as well as its limitations. 
Finally, Section \ref{sec:summary} summarizes this work.

\section{Kepler Observations} \label{sec:obs}

The \textit{Kepler} mission observed an specific area of the sky centered at R.A.$= 19^h22^m40^s$ and Dec$= +44^{\circ}30^{\prime}00^{\prime\prime}$ (J2000) between 2009 and 2013.
The observations were split into quarters and taken in two cadence mode, a short 1-minute cadence and a long 30-minute cadence.
The telescope camera is an array of 21 modules with two CCDs each. A CCD has two read-out channels adding to a total of 84 CCD channels. \textit{Kepler} telescope has an aperture of 0.95 meters, a 105 square deg field-of-view, and the instrument has a pixel scale of 3.98 arcsec per pixel.

Typically three FFIs were downlinked during quarter 5 and 17, one at the beginning, middle, and end of the quarter. A single FFI channel has $1100 \times 1024$ pixels, adding to a total of 95 mega pixels for the entire focal plane. Each FFI has an exposure time of 27 minutes.
\textit{Kepler}'s FFIs can be downloaded from the MAST archive\footnote{\url{https://archive.stsci.edu/missions-and-data/kepler/kepler-bulk-downloads}}.

\begin{figure}[htb!]
    \centering
    \plotone{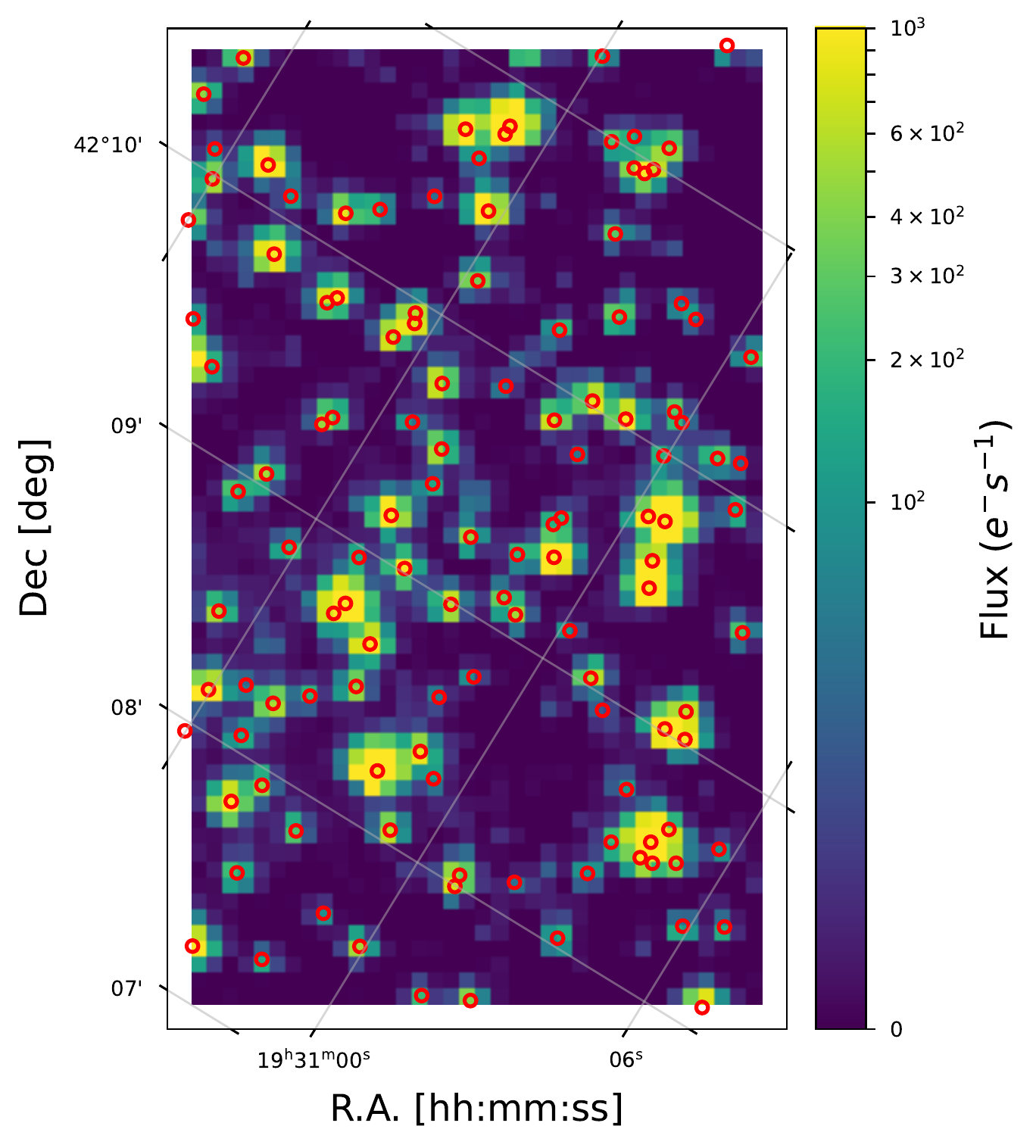}
    \caption{Example full EXBA mask (four tiles) for channel 48 observed during quarter 5. The image shown here is 36 x 60 pixels and the flux values averaged across cadences. 123 sources obtained from the Gaia EDR3 catalog down to $G \leq 20$ magnitude are marked with red circles (see Section \ref{subsec:exba_s}). The brightest visible star is  $G=13.7$ magnitudes. Our pipeline filters out all contaminating sources within $2 \arcsec$ to fit the PRF models. Sources outside the coverage of the mask (up to $4 \arcsec$) are allowed and extracted for photometry.}
    \label{fig:exba_img}
\end{figure}

The EXBA images were observed continuously between quarters 5 and 17 of \textit{Kepler}'s prime mission using the long cadence setup. 
These data consist in 4 custom mask of $9 \times 60$ pixels per channel, each of these 4 tiles are located next to each other in the CCD producing a $36 \times 60$ pixel image per channel. 
\textit{Kepler} EXBA images can be downloaded from the MAST archive\footnote{\url{https://archive.stsci.edu/kepler/}}, by searching with the \texttt{EXBA} keyword in the \textit{Investigation ID} field.
An example of the tiled EXBA mask for channel 48 is shown in Figure \ref{fig:exba_img}. 
Valid data exist for 80 out of 84 channels. The four channel are missing because of failure of circuitry powering in module 3 of the camera \citep{2016ksci.rept....2V}.

\section{Data Processing} \label{sec:data}

To obtain light curves based on aperture photometry that meet similar standards as those provided by \textit{Kepler}'s pipeline in the main target LCFs, we follow these steps:

\begin{enumerate}
    \item Create a detailed model of the PRF profile for each channel and quarter combination using \textit{Kepler}'s FFIs.
    \item Use Gaia catalogs to obtain accurate positions of the sources observed in the EXBA mask with a limiting magnitude of $G = 20$.
    \item Obtain a flux-normalized model of each source by evaluating the PRF model on all Gaia sources observed in the EXBA mask.
    \item Define multiple aperture masks according to the flux normalized source models and compute flux metrics that characterize them.
    \item Finally, compute multi-aperture photometry for every source, then compile and save the flux time series in FITS files following a similar structure as the LCFs.
\end{enumerate}

The following sections describe these steps in detail.

\subsection{Modeling the PRF Profile} \label{subsec:prf_ffi}

 \textit{Kepler}'s FFIs contain more than 1.1 million sources in total, on average $\sim 12,000$ sources can be detected in each channel. The large amount of sources available enables the creation of detailed PRF models for every channel.

\begin{figure*}[ht!]
    \centering
    \plotone{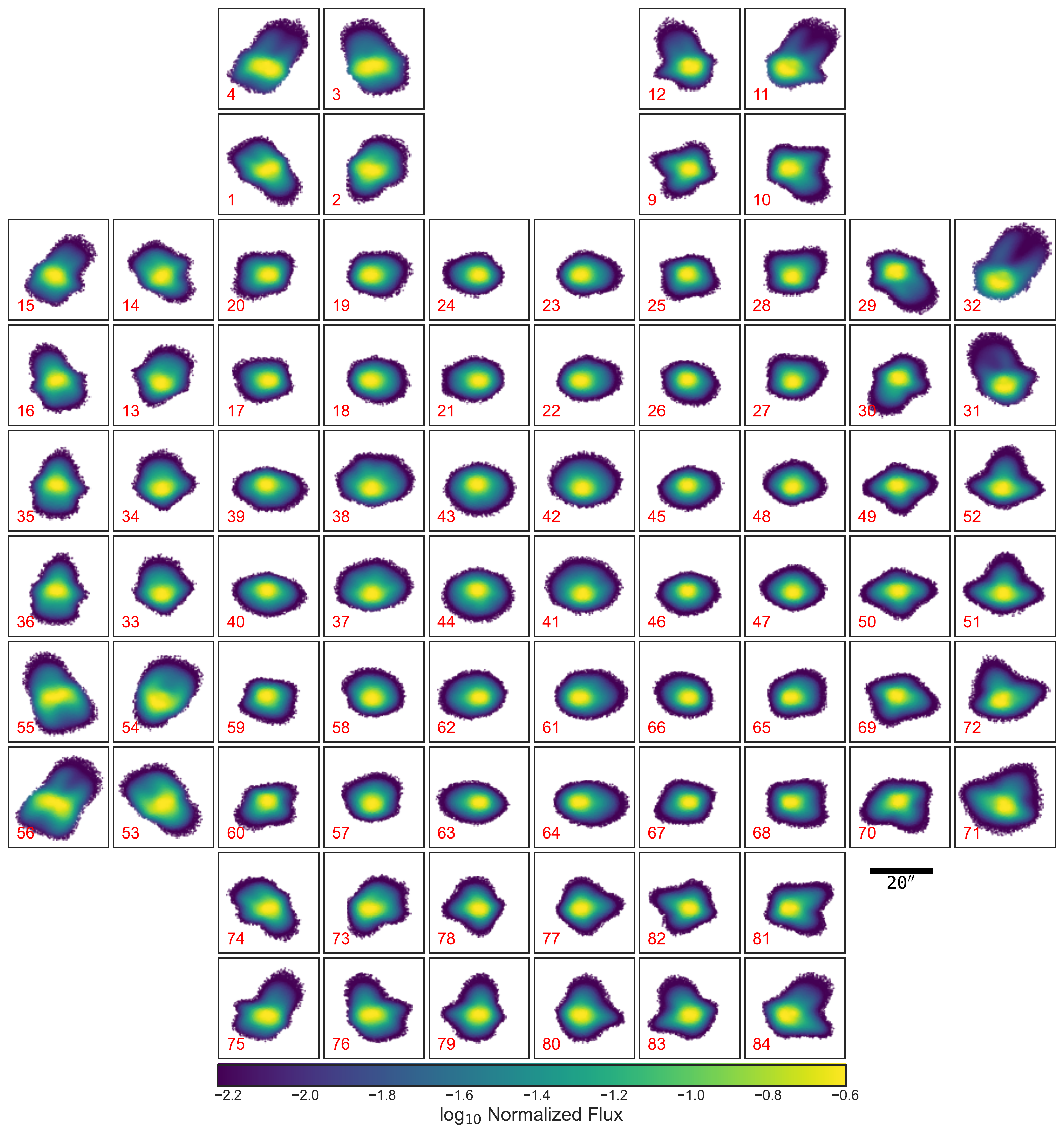}
    \caption{PRF models across the focal plane computed from \textit{Kepler}'s FFIs as described in \ref{subsec:prf_ffi}. The models are based on data observed during quarter 5. Each cell in the figure grid represents a single CCD channel (number in red). In a cell, each data point represents the pixel position from the reference source in Cartesian coordinates, while the color represents the pixel flux value normalized by the total flux of the corresponding reference source. 
    As discussed in \cite{2021AJ....162..107H}, the PRF profiles show steep gradients and are elongated, particularly as we move away from the center of the focal plane.
    Note that in this figure the CCD channels have the axis origin at each cell's lower-left corner, then the orientations of the plotted PRFs do not reflect the axis origin of the CCD channels on \textit{Kepler}'s focal plane which are arranged to be invariant under rotation. This is only aesthetics and does not affect the use of our models.
    }
    \label{fig:prf_ffi}
\end{figure*}

To compute PRF models of each channel, we use the PRF scene modeling described in the LFD photometry method \citep{2021AJ....162..107H}. 
This LFD photometry method uses Gaia catalogs to find and fix the coordinate position of sources in the image and models the scene PRF profile by fitting a linear model with spline basis to the data in polar coordinates.
The PRF modeling done by the LFD method assumes the following: it relies upon the accurate astrometry done by Gaia catalogs; the PRF shapes do not vary significantly over the spatial scale of the detector channel; and the PRF shapes do not change in time within each quarter.

Figure \ref{fig:prf_ffi} shows the PRF models for all 80 channels with available FFI data during quarter 5. 
At the center of the focal plane, the PRF shape is fairly circular, smooth, and shows angular symmetry. As we move away from the center, the PRF profile starts gradually showing elongation in both radial and tangential axis, this effect becomes extreme in the channels at the border. 
These characteristics of the data are conveniently captured by the model thanks to the use of polar coordinates, where the radial symmetry benefits the fit of a smoother model.

In \cite{2021AJ....162..107H}, the PRF profile is modeled using $~\sim 500$ Gaia sources and around $7,000$ pixel data points. The LFD method only uses pixel where flux from the sources is detected and removes all background pixels.
Here, in contrast, thanks to the large number of sources ($\sim 12,000$) in each FFIs channel image, the number of pixels reach a larger value $\sim 100,000$, allowing for more dense and robust representation of the PRF.
Additionally, \cite{2021AJ....162..107H} models the drift in position of all sources due to velocity aberration, this correction is only necessary when doing LFD photometry on all cadences of a TPF stack to obtain light curves.
In this work, we only follow the LFD method to obtain the PRF models from the FFIs and fitting the scene motion it is not only limited by the low number of FFI cadences but also out of the scope of this use-case.
Computing the PRF models from \textit{Kepler} FFIs can be computationally expensive, fortunately, the convenience use of \texttt{scipy}'s sparse matrices \citep{2020SciPy-NMeth}, keeps the memory use low ($\sim 450 MB$ for a single PRF model) and computing time in the order of $100$ seconds when using a modern personal laptop\footnote{2020 MacbookPro, 16 GB RAM, 2 GHz Quad-Core Intel Core i5 processor.}. Evaluation of a PRF model is faster, for the same machine it takes $\sim 1$ milliseconds per source.

Our pipeline produce a total of $1,080$ PRF models, one for every channel and quarter combination where the EXBA masks were collected.
These PRF models can be later evaluated for a grid of positions (in Cartesian coordinates) to obtain a flux normalized model for either a single source or a set of sources in any \textit{Kepler}-like data, e.g. TPFs, FFI, and SuperStamps\footnote{\citet{2018RNAAS...2Q..25C} created SuperStamp FITS files for K2 observation of open clusters (M35, M67, Ruprecht 147, and NGC 6530) and the Galactic bulge.} files.
Modeling the PRFs for every quarter capture the natural time degradation in sensitivity of the CCDs. The later is a factor that could have not been taken into account in \citet{2010ApJ...713L..97B} works due to the data availability, at the time of their work, only commissioning data was available. Furthermore, our PRF models use a small number of linear components which can be evaluated faster than grid-based models and are forced to be smooth with no edge effects. We direct the reader to \citet{2021AJ....162..107H} for further discussion.

\subsection{The EXBA Sources} \label{subsec:exba_s}

\begin{figure}[htb!]
    \centering
    \plotone{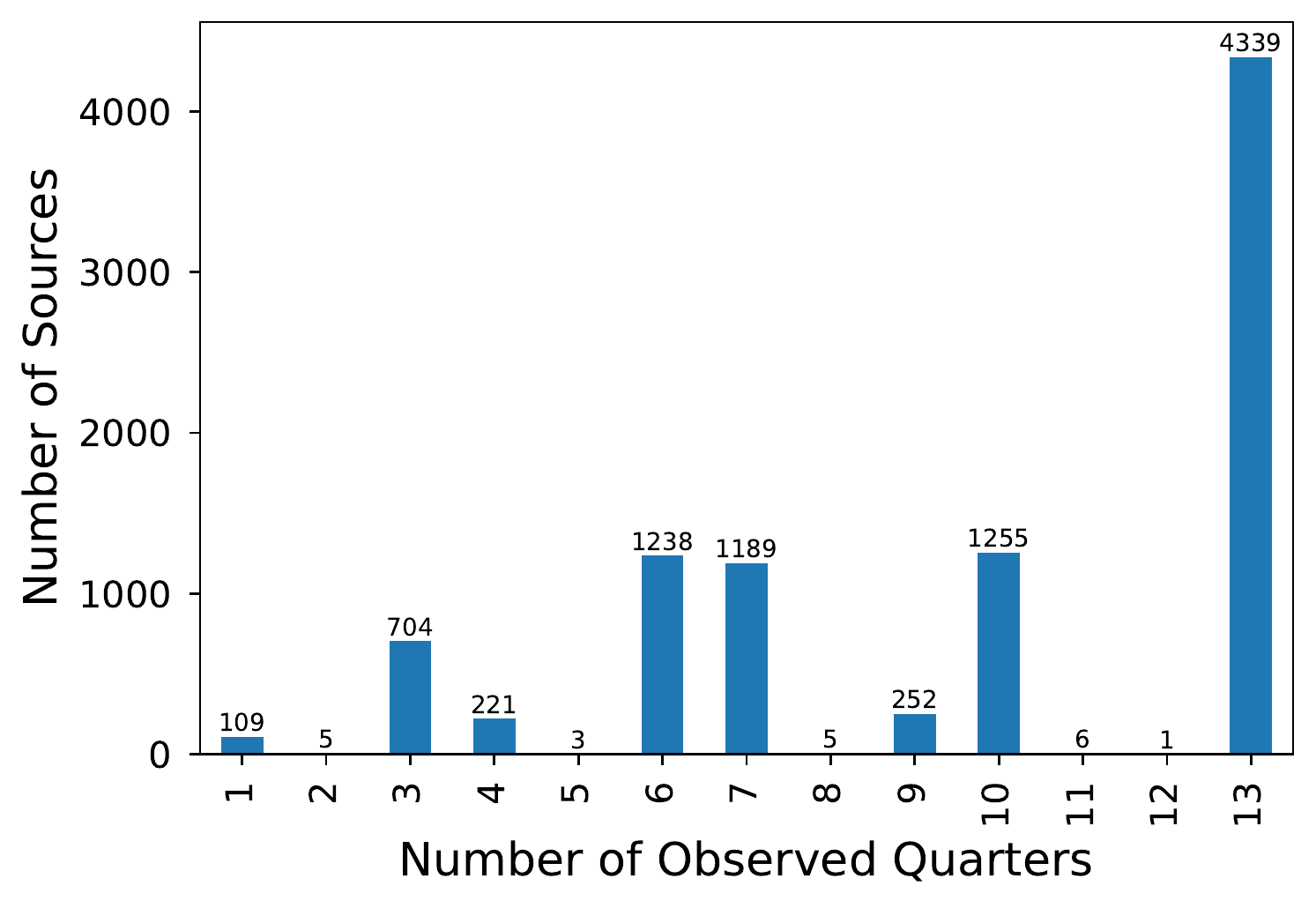}
    \caption{The number of EXBA sources with the total number of observed quarters. $46.5 \%$ of the total sources have observations in all 13 quarters that \textit{Kepler} used the EXBA masks. The rest have 12 or fewer quarters available.}
    \label{fig:source_hist}
\end{figure}

We use Gaia EDR3 \citep{2021A&A...649A...1G} as source catalog. 
This catalog provides high precision astrometry and source completeness, $0.5$ mas at $G = 20$ and, complete between $G = 12$ and $G = 17$, respectively \citep{2021A&A...649A...5F}. 
Although Gaia $G$ band \citep{2010A&A...523A..48J} does not exactly matches \textit{Kepler}'s response curve \citep{2016ksci.rept....1V}, it present an appropriate proxy for \textit{Kepler} magnitudes and can be used as a observed source catalog.
We query the Gaia catalog with a limiting magnitude of $G = 20$ and account for the source proper motions between each \textit{Kepler}'s quarter and the Gaia EDR3 epoch.
We flag contaminated sources ($< 5\%$) in crowded regions where the separation between sources is $< 2 \arcsec$ (0.5 of a pixel), and we allow for sources outside the limits of the EXBA mask up to $4 \arcsec$ (1 pixel). 
Only non-flagged sources will be extracted, but contaminants are considered when computing flux metrics (see following section for details).
After this, a total of $9,327$ (non-flagged) sources are found in the sky area covered by the EXBA mask ($\sim 0.2$ sq. deg.) across all quarters. 
A CCD channel covers the same sky area every four quarters due to the rotation of the spacecraft between quarters.
This effect, combined with the mask alignment in the rotational symmetric channels, leads to sources near the borders of the mask (within $20 \arcsec$) to fall outside of it in certain quarters. 
As a result, $45 \%$ of the sources have observations in all 13 quarters while $13\%$ of the sources have data in either 6, 7, or 11 quarters (see Figure \ref{fig:source_hist} for details).

\begin{figure}[htb!]
    \centering
    \plotone{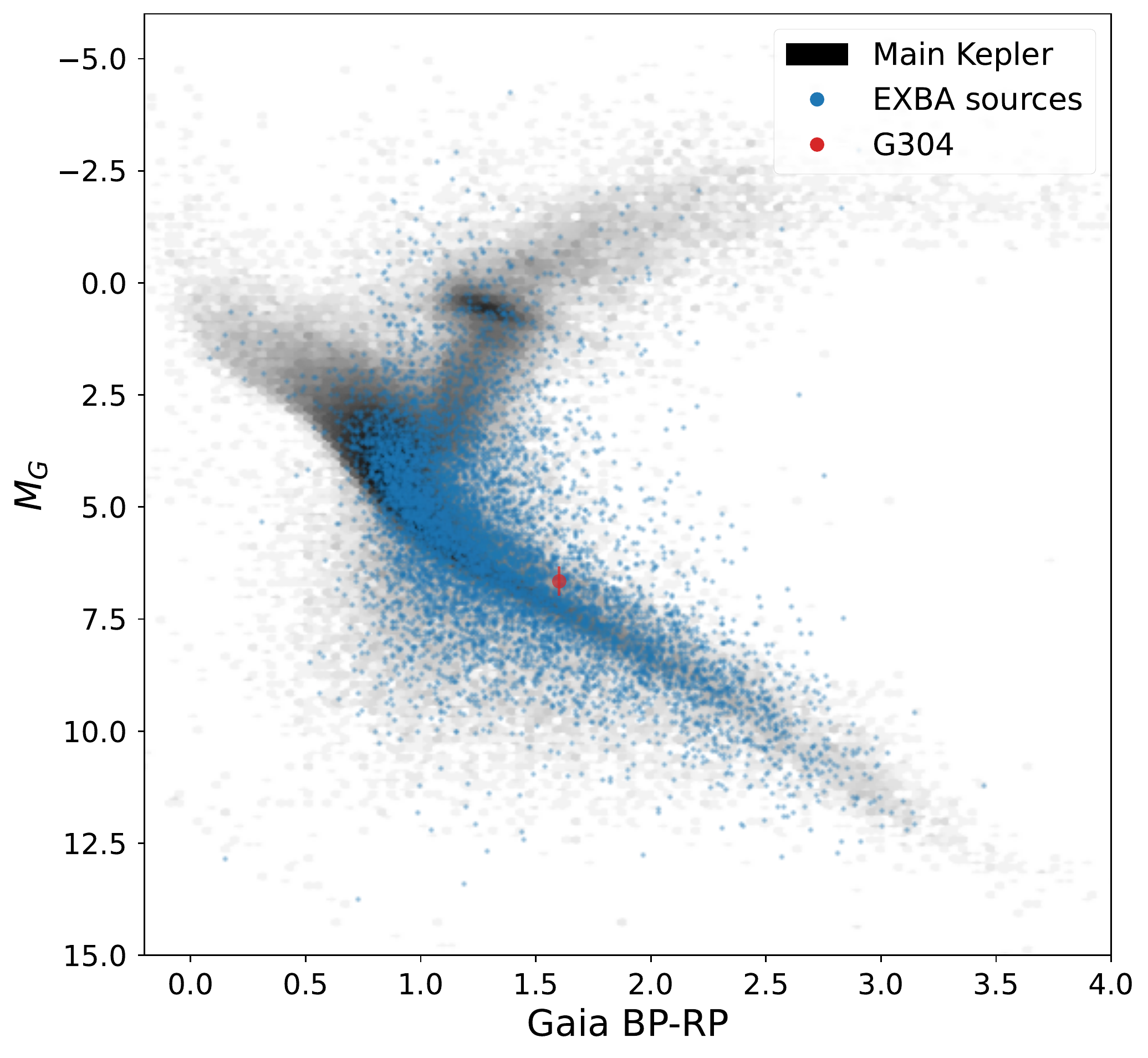}
    \caption{Color Magnitude Diagram of the EXBA sources (9,327 sources in blue) in comparison with the main targets (213,393 sources) of \textit{Kepler}'s mission (density plot in black). The red marker represents the location of \candidate (see Section \ref{subsubsec:g304}) that falls in the binary sequence.}
    \label{fig:cmd_exba}
\end{figure}

Figure \ref{fig:cmd_exba} shows the Color Magnitude Diagram (CMD) for both, main \textit{Kepler} Targets and the EXBA sources. 
The object photometry comes from Gaia EDR3 catalog and their distances computed by \cite{2021AJ....161..147B}.
The EXBA sources follow the distribution of \textit{Kepler}'s main targets except for the horizontal and asymptotic branches. A large fraction of the EXBA sources are located in the main sequence and binary sequence regions while a small fraction are evolved stars. The EXBA masks were chosen to be representative of the background field in comparison with the selection bias of \textit{Kepler}'s mission that specifically targeted evolved stars \citep{2010ApJ...713L.109B}.

\subsection{Aperture Photometry} \label{subsec:aper}

With the detailed PRF models computed from the FFI data for every channel and quarter combination we are able to evaluate them for every source in the EXBA images. 
We evaluate the PRF model on the Cartesian grid of pixel coordinates of the image data and with a maximum distance up to six pixels from the center of the source (obtained from the Gaia astrometry).
The PRF model evaluation leads to a normalized (to sum 1) flux model of the source.
In this work we evaluate the PRF model to obtain a normalized model of the sources aiming to capture the characteristic shape of the PRF, in contrast with the LFD photometry method where the evaluation aims to obtain the total flux of the source.

Following \emph{Kepler}'s crowding and flux loss \citep[CROWDSAP and FLFRCSAP respectively, ][]{2012PASP..124..963K} metrics that the pipeline calculated for the photometric aperture of every primary target, we compute two equivalent metrics. 
Per definition, FLFRCSAP is the fraction of target flux contained in the photometric aperture over the total target flux. 
While CROWDSAP is the ratio of target flux relative to the total flux within the photometric aperture including contaminating sources. 
Therefore a CROWDSAP value of 1 means no contamination and a FLFRCSAP value of 1 means the aperture contains the entire flux of the source.
Evaluating the PRF model for all sources and given their photometric aperture, we can compute both crowding and completeness metrics for every source.

We use isophotes to define the boundaries of the photometric apertures.
The boundary of each isophote for a given source is defined by the flux value at a certain percentile of the normalized flux distribution. The seven percentiles are $0, 15, 30, 45, 60, 75$ and $90$.
This leads to seven different apertures that follow the characteristic PRF shape and enclose different levels of flux completeness and contamination.
Per definition, larger percentile values lead to smaller apertures and therefore less contamination (larger CROWDSAP values) but less flux completeness (smaller FLFRCSAP values).
Additionally, we include an extra aperture in which the percentile that defines the isophote value is computed by simultaneously optimizing both metrics, FLFRCSAP and CROWDSAP. 
This scalar optimization is performed with \texttt{}{scipy}'s Brent's method\footnote{\url{https://docs.scipy.org/doc/scipy/reference/generated/scipy.optimize.minimize_scalar.html}}. 
After the target value is reached for either of the metrics, a softening term\footnote{We redefine the updated metric value as $metric^{\prime} = target + leakFactor(metric - target)$ where the $target$ is the desired target value for the metric (e.g. $0.9$), $metric$ is the current value, and $leakFactor$ is a user defined value (e.g. $0.001$).} is applied to allow the optimization to focus in the second metric until both reach their respective target scores, instead of driving a single metric to the optimal point. 
We optimize the aperture size by targeting for a CROWDSAP $> 0.8$ and a FLFRCSAP $>0.5$.

Providing multiple aperture photometry allows the users to choose the most suitable one depending on their scientific goal. 
Each aperture photometry has its associated FLFRCSAP and CROWDSAP metrics.
Later, as indicated in Kepler Archive Manual \citep{2016ksci.rept....9T}, crowding corrected fluxes can be computed by applying the following equation:

\begin{equation}
    \label{eq:crwd_eq}
    F_{corrected} = F - med(F)(1 - \rm{CROWDSAP})
\end{equation}

where $F$ is the original flux as provided in the light curve files, CROWDSAP is the crowding metric, and $med(F)$ is the median value of the uncorrected fluxes.
Finally, both metrics can be used to stitch and adjust the light curves from different quarters to account for aperture variations due to changes in the PRF profile.

\section{Results} \label{sec:results}

\subsection{Data Products} \label{subsec:data_products}

In this work we produce two high-level data products. 
First, a data set of light curves with multiple aperture photometry for $9,327$ sources observed by \textit{Kepler} in the EXBA masks. We defined the multi-apertures as seven different isophote levels of the PRF model evaluated at every source.
Secondly, a set of \texttt{Python} tools that allow the use of our precomputed PRF models to perform aperture photometry in any \textit{Kepler}-like data, such as TPFs and the EXBA data. 

The light curves provided in this work are stored at the MAST archive\footnote{Kepler Bonus, APEXBA,  {\doi{10.17909/t9-d5wy-e535}}} and can be accessed similarly as the main \textit{Kepler}'s LCFs. 
Each FITS file contains light curves of a single EXBA source observed during a quarter. 
The FITS files are multi-extension, with a table containing the flux time series from multi-aperture photometry, and 8 extra image extension containing the aperture mask used for every photometric aperture. 
There are a total of $90,224$ LCFs, which accounts for the number of observed sources and the number of quarters each source was observed.
Each file header contain relevant keywords such as Gaia's source information, EXBA origin file, and the completeness and contamination metrics for all the photometric apertures. 
Alongside the LCFs, a catalog with all observed Gaia sources for each channel and quarter is also available. 
Appendix \ref{appx:lcfs} details the structure and content of each EXBA LCF.

Accompanying this dataset of light curves we release \texttt{kepler-apertures}\footnote{\url{https://github.com/jorgemarpa/kepler-apertures/releases/tag/v0.1.0}}, a \texttt{Python} library that allows to reproduce the results presented in this work.
The PRF models computed from \textit{Kepler}'s FFIs are included in the library.
Through using the this PRF models users can build aperture masks and compute flux metrics for any type of \textit{Kepler}'s prime mission data. 
\texttt{kepler-apertures} also provides the functionalities to compute new PRF models from FFI data by using the LFD method.

\subsection{Finding Transiting Signals} \label{subsec:transit_search}

As planned, the \textit{Kepler} pipeline only produced TPFs for the EXBA mask and no light curves were extracted. Thus, the EXBA data has never been systematically analyzed to search for exoplanet candidates, or any other time domain studies. 
We follow a standard approach to search for transiting signals.

\subsubsection{Detrending} \label{subsubsec:detrending}

Before analyzing each light curve for scientific purposes, we process them to correct for instrumental effects. 
Not only that, numerous stars show stellar activity due to rotation or pulsation, these signals need to be removed to isolate small variations of the incoming flux originated from transit events.
Removing these types of signals is often refer as detrending.

We use the Cotrending Basis Vectors \citep[CBVs, ][]{2012PASP..124.1000S} to account for the systematic signals introduced by the instrument.
CBVs are a series of 16 basis vectors that capture most of the known correlated features in a reference ensemble of flux time series. 
The \textit{Kepler} pipeline computed these basis vectors for every quarter and output channel combination.
As shown in \cite{2012PASP..124.1000S}, the first 8 CBV components capture most of the systematic errors, therefore for this work we only use the first eight basis vectors.
To capture variability due to stellar activity, such as rotation or pulsation, we use a set of basis splines. 
Similar to \cite{2014PASP..126..948V} we found that the number of knots for the spline basis equivalent to a spacing of 1 day allows fitting variability in short time scales without losing transit-like signals.
Combining the aforementioned set of basis vectors we defined a design matrix that is fitted using linear regression to the data and obtain the corrected time series. 
We perform this detrending correction for all the source light curves on a per quarter basis.
As an example, Figure \ref{fig:g304_lc} shows the flux time series of \candidate. 
The light curve corrected only with the CBVs (upper panel) shows clear stellar variability in the scale of days. 
The light curve after full detrending (middle panel) preserves the transiting signal. 
Although the fully corrected light curve still shows residual stellar variability (as seen in the 1 day binned light curve shown in black), the amplitude is significantly smaller that the transiting signal.

\subsubsection{Box Least Square Periodogram} \label{subsubsec:bls}

We perform a search for periodic transiting signals using the Box Least Square Periodogram \citep[BLS,][]{2002A&A...391..369K}. 
The BLS method searches for periodic variability by fitting the data with an upside-down top hat periodic model. 
The model is parameterized by the period, transit duration, and the time at mid-transit.

Due to the structure of our light curves, which are separated by quarters, we compute the BLS periodograms in two time regimes. 
First, on a per quarter basis search using a grid periods between 0.2 and 30 days. 
The grid consist in $60,000$ values evenly spaced in frequency.
The upper limit allows for detecting at least three transits within the $\sim 90$ days time window of each quarter. 
Secondly, we search for signals with periods ranging from 60 to 360 days, for this we use the object light curves from all available quarters.
The BLS periodogram return us the parameters of the model that best fits the data.

To remove periodic signals originated from high-amplitude eclipsing binary systems and other types of periodic pulsators we filtered the BLS parameters results. 
This filtering isolates transiting signal potentially originated from small companions. 
We only consider sources with the following BLS results: 
1) transit depth lower than $10 \%$ in normalized flux units, 
2) a signal-to-noise ratio of the periodogram maximum power larger than $20$, and 
3) a periodogram maximum power larger than $20$ after standardizing the periodogram by the mean and standard deviation of the ensemble of periodograms. 

\begin{figure}[htb!]
    \centering
    \plotone{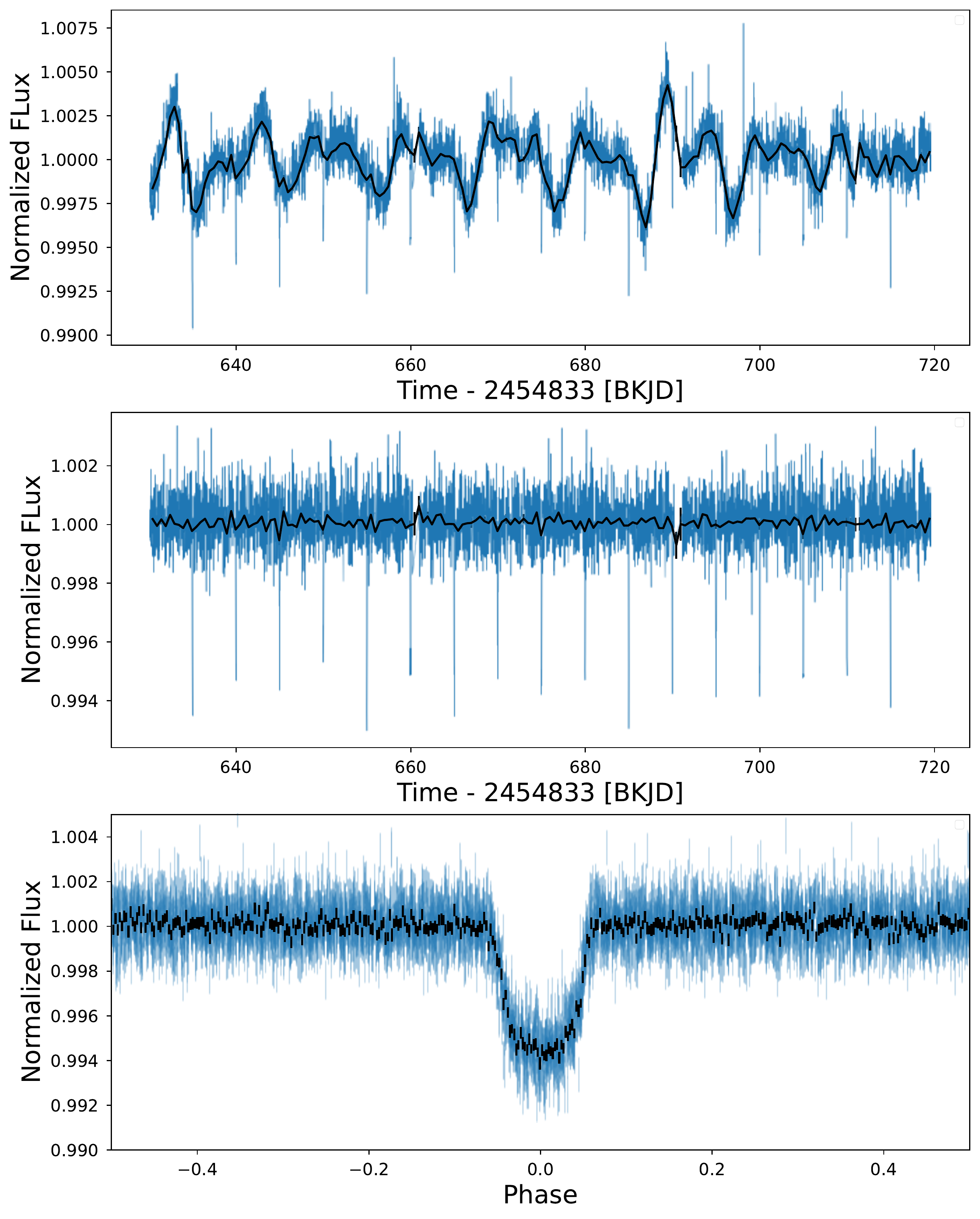}
    \caption{Time-based light curve (top two panels) and phase-folded light curve (bottom panel) of \candidate. Blue markers are the flux measurements, while black markers are the binned light curves using a time bin size of 1 day for the time light curve. The top and middle panels show the CBV-only corrected and fully detrended (CBVs and splines) light curves both respectively, for observations during quarter 7. The phase-folded light curve contains data from all quarters as described in Section \ref{subsubsec:g304}. The transit u-like shape and depth ($\sim 0.5 \%$) evidence the possibility of a large planet candidate or a sub-stellar companion. The parameters that characterize the best transit model are listed in Table \ref{tab:G304_params}.}
    \label{fig:g304_lc}
\end{figure}

After a visual inspection of less than two hundred light curves after the aforementioned filtering, one source stands out. 
\candidate (here and after \textit{G...304}) exhibits the characteristic transit shape and depth of a large planet or a sub-stellar companion. 
Figure \ref{fig:g304_lc} shows both time-based (quarter 7) and phase folded light curve of \textit{G...304}.

\subsubsection{\candidate} \label{subsubsec:g304}

\begin{figure}[htb!]
    \centering
    \plotone{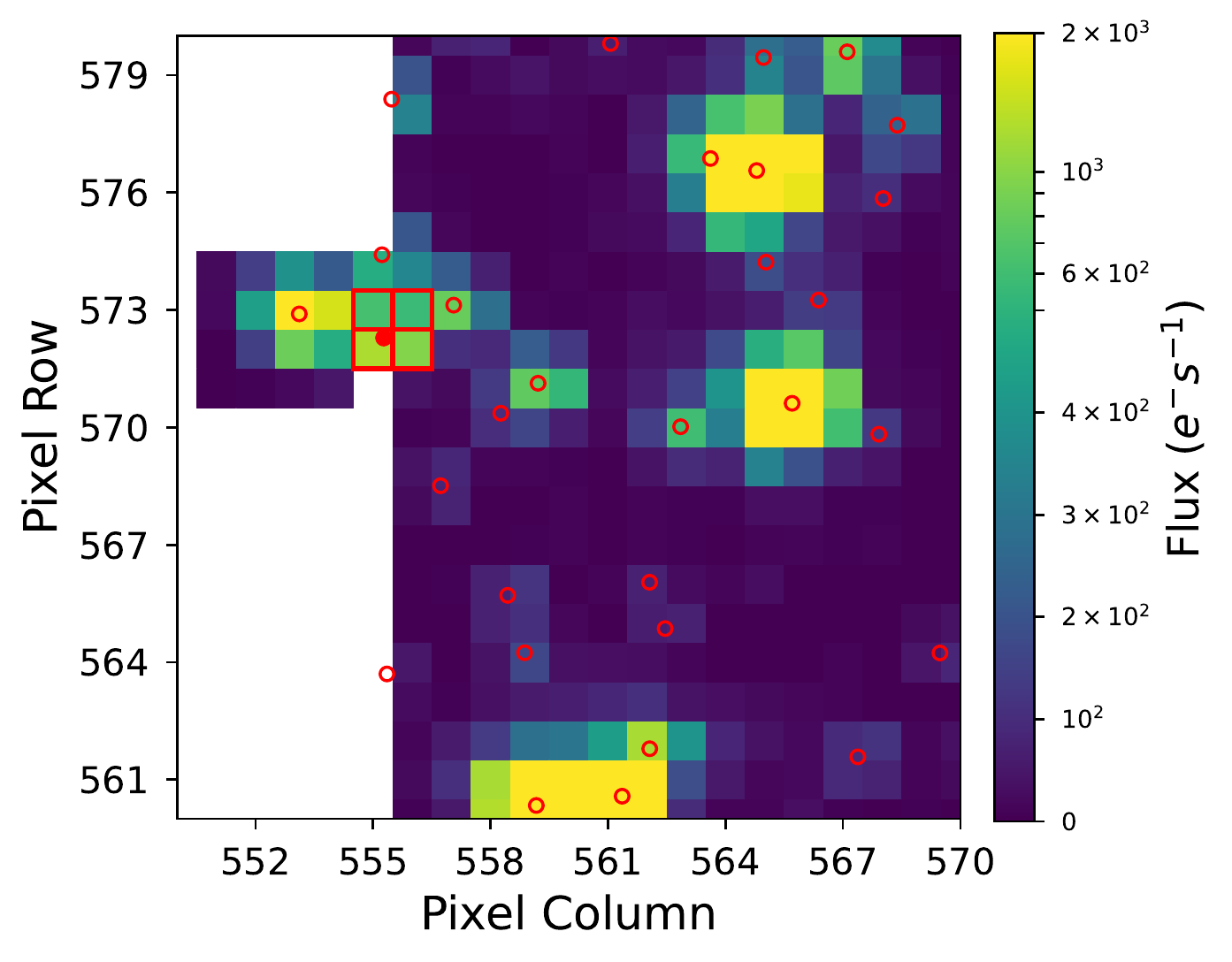}
    \caption{Image in pixel space of the EXBA data ($\rm{Pixel\ Column} \geq 556$) and the KIC 7214804 TPF ($\rm{Pixel\ Column} < 556$) where the host candidate \candidate is located. The flux data is an average of cadences observed during quarter 7. Red circles show Gaia EDR3 sources, while the solid circle locates the host candidate. The red squares illustrate the aperture mask used for this work.}
    \label{fig:g304_img}
\end{figure}

\textit{G...304} was observed in the EXBA masks during quarters 7, 8, 11, 12, 15, and 16, and it is located outside the coverage of the mask (within 1 pixel from the mask border) according to Gaia astrometry. 
Moreover, it was also observed as a background source in the main TPF of KIC 7214804 continuously between quarters 2 and 17.
As shown in Figure \ref{fig:g304_img}, Gaia astrometry (solid red dot in the figure) locates \textit{G...304} inside the TPF, but sufficient flux is contained in the EXBA data to detect the transit events.

Because the EXBA masks did not capture \textit{G...304} during quarters 2 through 6, 9, 10, 13, 14, and 17, we decided to work only with quarters that have both EXBA and TPF data.
Combining the EXBA and TPF pixel data we are able to create a "stitched" pixel image which improves flux completeness leading to a more accurate photometry.
Not using all available quarters does not impact the number of observed transits in a significant amount and allows to obtain accurate model fitting. 
We compute a new photometric aperture from the combined image that maximizes the transiting signal using the \texttt{contaminante}\footnote{\url{https://github.com/christinahedges/contaminante}} \citep{2021RNAAS...5..260H} \texttt{Python} package.
Given a period value and transit reference time, \texttt{contaminante} creates an aperture mask by finding which pixels contain the transiting signal at high significance, marginalizing over known instrument systematics.

We defined the orbital parameters of \textit{G...304} using the BLS algorithm at a finer period grid. We found a transit period of $4.99667$ days, a transit reference time of $2455468.058$ Julian days, a duration of $0.091$ days, and a normalized transit depth of $0.005$. 
An odd-even transit depth test \citep{2010SPIE.7740E..19W} helps to separate transit signals originated from eclipsing binaries.
We perform this test with the best BLS model parameters resulting in a test statistic value of $0.00235$ and significance level of $96.17\%$  indicating the odd and even transits are statistically consistent.
The transiting signal is a viable case for a transiting planet (see Figure \ref{fig:g304_lc}, lower panel), with a transit-like shape and a transit depth consistent with a large planet or sub-stellar companion. 
\textit{G...304} light curve also evidences significant variability due to stellar activity in the scale of days.

Unfortunately, Gaia DR2 and EDR3 do not provide estimates of the stellar radius, luminosity, or line-of-sight extinction for \textit{G...304}, only an estimate of the effective temperature ($T_{eff} = 4,358$ K). 
To constrain the host star parameters and understand the evolutionary state of \textit{G...304}, we perform a spectral energy distribution (SED) analysis of the available Gaia ($G$, $B_P$, and $R_P$, \citep{2018A&A...616A...1G}), 2MASS ($J_{2\rm{M}}, H_{2\rm{M}}$, and $K_{2\rm{M}}$ \citep{2003yCat.2246....0C}), and WISE (W1 and W2, \citealp{2012yCat.2311....0C}) photometry using the publicly available exoplanet fitting suite, \texttt{EXOFASTv2} \citep{2013PASP..125...83E, 2019arXiv190709480E}.
We included the MESA Isochrones and Stellar Tracks (MIST) stellar evolution models \citep{2011ApJS..192....3P, 2013ApJS..208....4P, 2015ApJS..220...15P, 2016ApJ...823..102C, 2016ApJS..222....8D} within the fit to precisely estimate the host star parameters. 
A Gaussian prior on the parallax from Gaia DR2 of 1.05010$\pm$0.22882 mas (corrected for the -30 $\mu$as offset reported by \citet{2018A&A...616A...2L} was enforced on the fit and we also include an upper limit of 0.6677 mag on the line-of-sight extinction (A$_V$) from \citet{1998ApJ...500..525S} \& \citet{2011ApJ...737..103S}. 
Lastly, we assume a solar metallicity prior of 0.0$\pm$0.25 dex. 
Our analysis showed that \textit{G...304} is a K-dwarf star with a mass of $0.749 M_{\odot}$, a stellar radius of $0.720 R_{\odot}$, effective temperature of $4,410$ K consistent with the value provided by Gaia DR2, and a density of $2.83 g/cm^3$.
Table \ref{tab:G304_params} summarizes these results.

We model the light curve of \textit{G...304} using the \texttt{exoplanet} library \citep{exoplanet:exoplanet} and \texttt{pymc3} \citep{10.7717/peerj-cs.55}.
These tools allow to set priors for the stellar parameters, such as the stellar radius (R$_*$), density ($\rho_*$) and Limb Darkening, and for the orbital parameters such as period ($P$), transit reference time ($t_0$), impact parameter ($b$) and the ratio between the planet and the star (R$_p / $ R$_*$).
In a similar manner as \cite{Sandford_2017} and including the stellar parameter results from the SED fitting for this system, we adopt the following priors:

\begin{itemize}[label={\tiny\raisebox{1ex}{\textbullet}}]
    \item $t_0$: a uniform prior bounded by $t_{0\rm{, BLS}} - 0.5$ days and $t_{0\rm{, BLS}} + 0.5$, where $t_{0\rm{, BLS}}$ is the transit time at maximum power from the BLS periodogram.
    \item $P$: a uniform prior from $0.9P_{\rm{BLS}}$ to $1.1P_{\rm{BLS}}$ days, where $P_{\rm{BLS}}$ is the period at maximum power from the BLS periodogram.
    \item $b$: a uniform prior from 0 to 2.
    \item R$_p / $ R$_*$: a uniform prior from 0 to 1.
    \item quadratic Limb Darkening coefficients, $u_1$ and $u_2$, following the uninformative prior defined by \cite{2013MNRAS.435.2152K}.
    \item $\rho_*$: reparameterized as $log_{10}(\rho_*[kg/ m^3])$ and a uniform prior in log space between 0 and 3.5, which is the stellar density from the SED fitting in units of $[kg/ m^3]$.
    \item R$_*$: a uniform prior between 0 and $1.1 \rm{R}_{*, \rm{SED}}$.
\end{itemize}

\textit{G...304} locates close to the binary sequence in the HR diagram (red marker in Figure \ref{fig:cmd_exba}) and has a large value of Gaia's Renormalized Unit Weigh Error (RUWE = 4.2)\footnote{\url{https://gea.esac.esa.int/archive/documentation/GDR2/Gaia_archive/chap_datamodel/sec_dm_main_tables/ssec_dm_ruwe.html}} suggesting that the target could be an unresolved binary system.
Considering the relation between the RUWE and the photocentre perturbation derived by \cite{10.1093/mnras/staa1522}, \textit{G...304} has a centroid perturbation of $\sim 2$ mas ($\sim 2$ AU at 970 pc distance).
Considering a physically possible binary systems with a K-dwarf primary (as the SED fitting suggested) and a K-dwarf (or a smaller M-dwarf) secondary with a 5-day period orbit and an estimated $0.065$ AU semi-major axis (assuming non-eccentric Keplerian orbit) can not account for a 0.5\% transit depth and the measured astrometric scatter.
Therefore, the transits must be produced by a third body orbiting one of the stars in the binary system.
We added a dilution term to our light curve model that accounts for the flux provided by a companion following the parametrization suggested by \citet{2019MNRAS.490.2262E}:

\begin{equation}
    \label{eq:dilution}
    \mathcal{M}(t) = [\mathcal{T}(t)D + (1 - D)] \left( \frac{1}{1 + DM}\right)
\end{equation}

where $\mathcal{T}(t)$ is the transit model defined by the stellar and orbital parameters, $D$ is the dilution term, and $M$ is the out-of-transit target flux.

\begin{figure*}[htb!]
    \centering
    \plotone{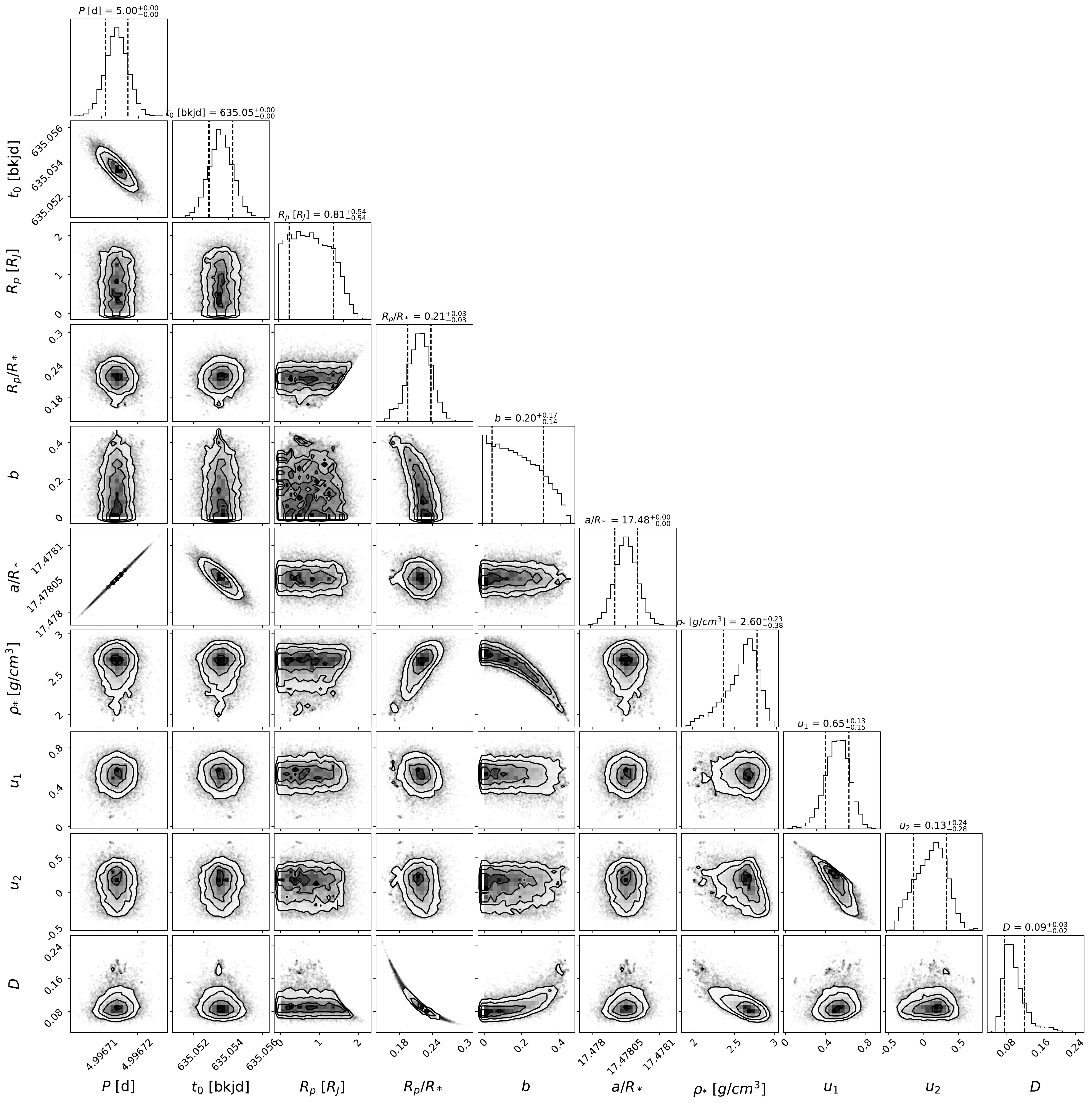}
    \caption{Posterior distributions of inferred stellar and transit parameters for the transiting signal observed around \candidate. The model fitting was performed using the \texttt{exoplanet} Python package following the priors described in Section \ref{subsubsec:g304}. The posteriors suggest the eclipse signal is produced by a large planet with a radius $< 2 R_J$.
    \label{fig:g304_corner}}
\end{figure*}

\begin{figure*}[htb]
    \centering
    \plotone{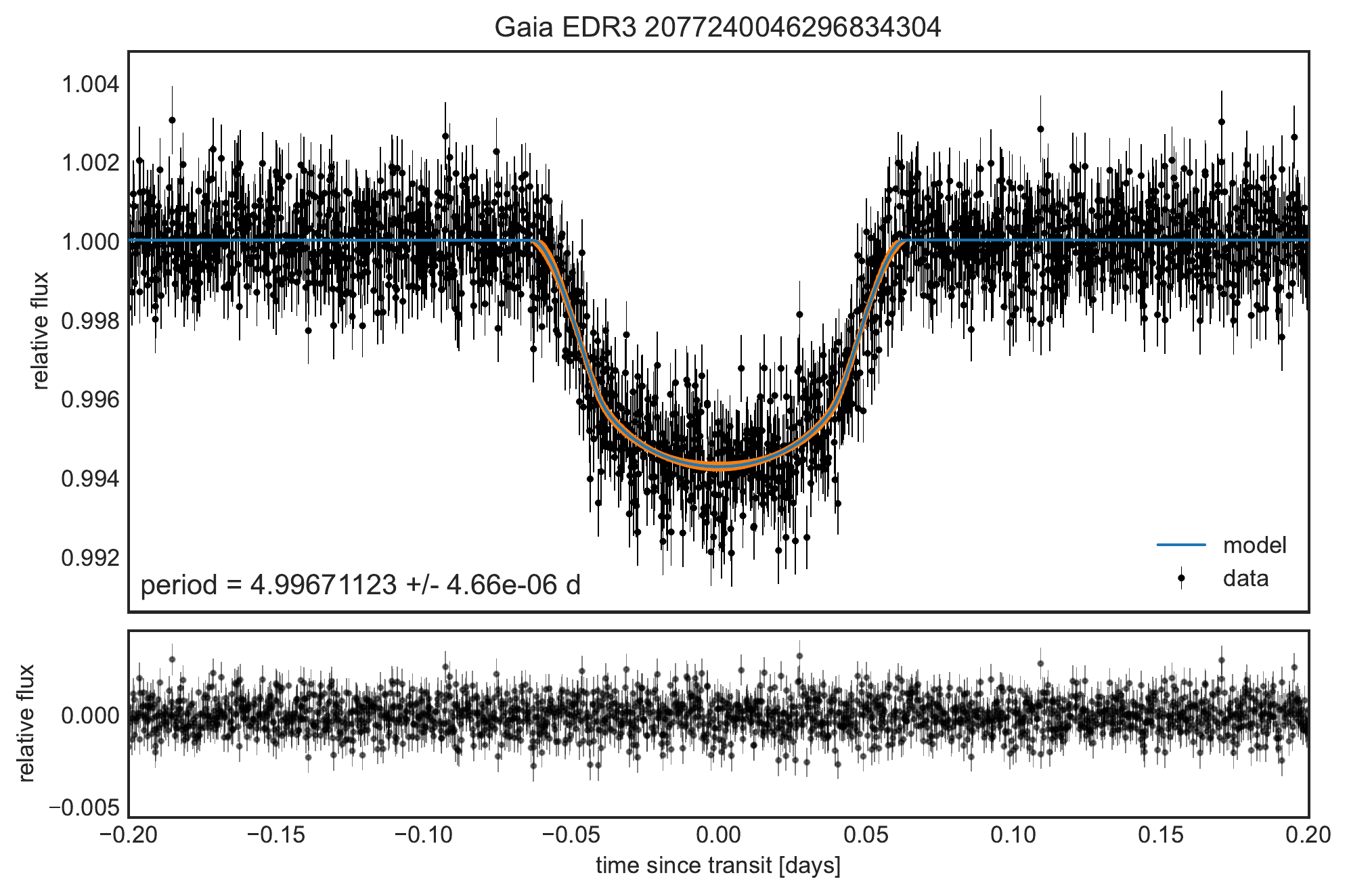}
    \caption{Transit model fitted using the \texttt{exoplanet} tool. The upper panel shows the light curve data (black markers), the median model (solid blue line), and the 3rd and 97th percentile (in orange) from the posterior distribution. Residuals are shown in the lower panel with no clear evidence of a remaining signal. The transit shape and depth are typical of a large planet companion.}
    \label{fig:g304_lc_fit}
\end{figure*}

The results of the model fitting are presented in Table \ref{tab:G304_params} while the joint posterior distributions are shown in Figure \ref{fig:g304_corner}. 
The inferred parameters for the radius $0.81 ^{+0.54}_{-0.54} R_{J}$ of the candidate (with a maximum radius of $2.2 R_{J}$ from the posterior distribution), indicate the transits are produced by a substellar companion orbiting at $0.034$ AU.
Subtracting the most representative model to the data shows no remaining signal, as shown in Figure \ref{fig:g304_lc_fit}.
The inferred stellar density ($2.60 g/cm^3$) is in agreement (within the errors) with the value obtained from the SED fitting. 
The mean value in the stellar parameters inferred from the model fitting ($R_{*} = 0.4 \pm 0.26 R_{\odot}$ and $M_{*} = 0.116^{+0.54}_{-0.11} M_{\odot}$) differ with those find from the SED fitting ($R_{*, \rm{SED}} = 0.720 ^{+0.048}_{-0.046} R_{\odot}$ and $M_{*, \rm{SED}} = 0.749 ^{+0.048}_{-0.049} M_{\odot}$) due to the inclusion of the dilution term. 
During the SED fitting, we treated the target as a single star. 
Including the dilution term also affected the width of the posterior distributions, enlarging the derived uncertainties for stellar radius and mass.

\begin{deluxetable*}{lll}[htb!]
\tablecaption{Best model parameters from the SED fitting (first block) and the transiting signal fitting using the \texttt{exoplanet} library (second and third block) around \candidate. The last block details the host Gaia EDR3 values.
\label{tab:G304_params}}
\tablewidth{0pt}
\tablehead{
\colhead{Parameter} & \colhead{Description} &\colhead{Value}
}
\startdata
    $R_{*, \rm{SED}}$ \dotfill    & Radius [$R_{\odot}$] \dotfill           & $0.720 ^{+0.048}_{-0.046}$ \\
    $M_{*, \rm{SED}}$ \dotfill    & Mass [$M_{\odot}$] \dotfill             & $0.749 ^{+0.048}_{-0.049}$ \\
    $L_{*}$ \dotfill              & Luminosity [$L_{\odot}$] \dotfill       & $0.175 ^{+0.044}_{-0.034}$\\
    $\rho_{*, \rm{SED}}$ \dotfill & Stellar density [$g / cm^3$] \dotfill   & $2.83 ^{+0.48}_{-0.40}$ \\
    $log (g)$ \dotfill            & Surface gravity [$cm / s^2$] \dotfill   & $4.598 \pm 0.04$ \\
    $T_{\rm{eff, SED}}$ \dotfill  & Effective Temperature, SED [K] \dotfill & $4410 ^{+150}_{-140}$ \\
    $[$Fe/H$]$ \dotfill           & Metallicity (dex) \dotfill              & $0.32 ^{+0.15}_{-0.18}$ \\
\hline
    $R_*$ \dotfill    & Radius [$R_{\odot}$] \dotfill         & $0.400 ^{+0.26}_{-0.26}$ \\
    $M_*$ \dotfill    & Mass [$M_{\odot}$] \dotfill           & $0.116 ^{+0.39}_{-0.11}$ \\
    $\rho_*$ \dotfill & Stellar density [$g / cm^3$] \dotfill & $2.60 ^{+0.23}_{-0.38}$ \\
    $u_1$ \dotfill    & Limb Darkening Coefficient 1 \dotfill & $0.65 ^{+0.13}_{-0.15}$ \\
    $u_2$ \dotfill    & Limb Darkening Coefficient 2 \dotfill & $0.13 ^{+0.24}_{-0.28}$ \\
    $D$ \dotfill      & Dilution \dotfill                     & $0.09 ^{+0.03}_{-0.02}$ \\
\hline
    $R_{P}$ \dotfill  & Radius [$R_{Jup}$] \dotfill      & $0.81 ^{+0.54}_{-0.54}$ \\
    $P$ \dotfill      & Period [days] \dotfill           & $4.99671123 \pm 4.66 \times 10^{-6}$ \\
    $t_{0}$  \dotfill & Transit Mid Point [BKJD] \dotfill & $635.0541 \pm 0.0005$ \\
    $t_{14}$ \dotfill & Duration [hours] \dotfill        & $0.1 \pm 1 \times 10^{-6}$\\
    $b$ \dotfill      & Impact parameter \dotfill        & $0.2 ^{+0.17}_{-014}$\\
    $a$ \dotfill      & Semi-Major Axis [AU] \dotfill    & $0.032 ^{+0.02}_{-0.02}$ \\
    $a/R_*$ \dotfill  & Semi-Major Axis / R$_*$ \dotfill & $17.48 \pm 1 \times 10^{-5}$ \\
\hline  
    $\alpha_{\rm{J2000}}$ \dotfill & R.A. \dotfill                       & $295.859489 \pm 0.000132$ \\
    $\delta_{\rm{J2000}}$ \dotfill & Decl.  \dotfill                     &  $42.713179 \pm 0.000142$ \\
    $\mu_\alpha$ \dotfill.         & Proper Motion RA [mas/yr]  \dotfill & $-1.99 \pm 0.17$ \\
    $\mu_\delta$ \dotfill.         & Proper Motion Dec [mas/yr] \dotfill & $-9.40 \pm 0.20$\\
    $\pi$ \dotfill                 & Parallax [mas/yr] \dotfill          & $1.03 \pm 0.15$\\
    $T_{\rm{eff, Gaia}}$ \dotfill  & Effective Temperature, Gaia DR2 [K] \dotfill & $4358.5 ^{+244.7}_{-351.5}$ \\
    G  \dotfill                    & Gaia EDR3 G magnitude \dotfill      & $16.5922$ \\
    $R_{\rm{P}}$ \dotfill          & Gaia EDR3 $R_{\rm{P}}$ magnitude \dotfill    & $15.6498$ \\
    $B_{\rm{P}}$ \dotfill          & Gaia EDR3 $B_{\rm{P}}$ magnitude \dotfill    & $17.2516$ \\
\enddata
\end{deluxetable*}

\subsection{Eclipsing Binaries} \label{subsec:ebs}

\begin{figure*}[htb!]
    \centering
    \plotone{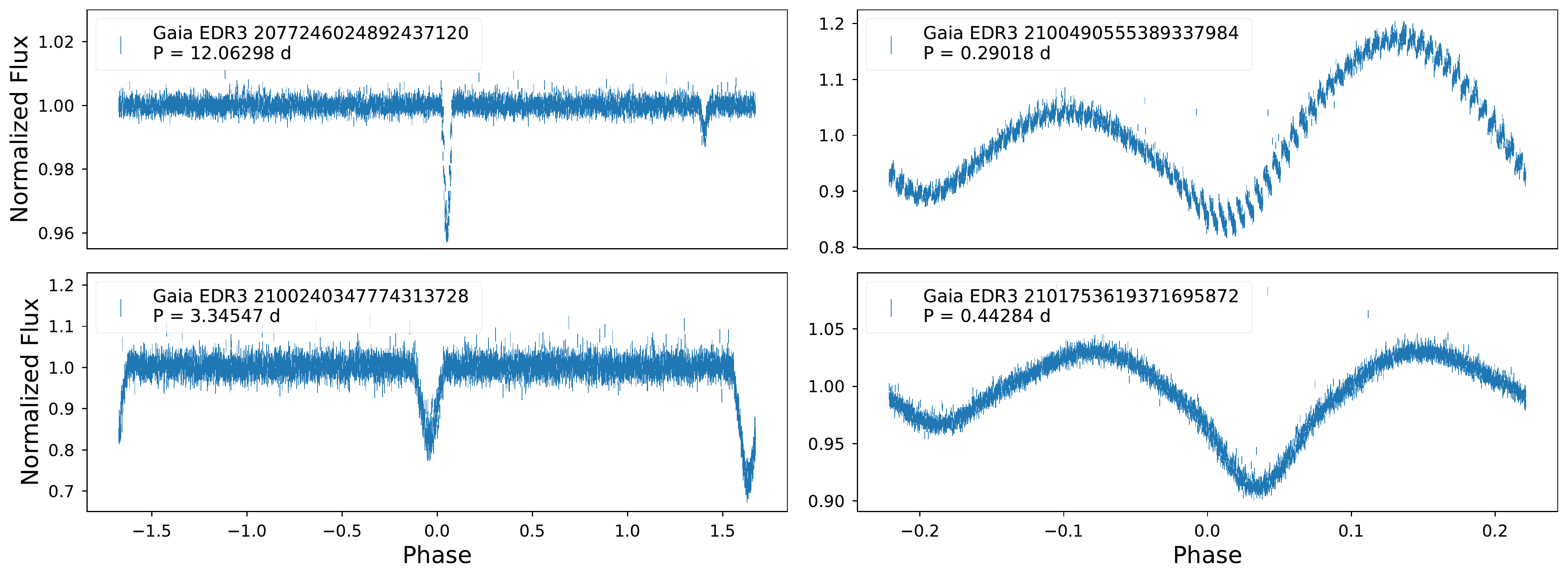}
    \caption{Examples of Eclipsing Binaries found in \textit{Kepler}'s EXBA masks. Detached systems (EAs) show well defined primary and secondary eclipses and are the major source of false positive when looking for exoplanet candidates. Contact systems (EWs) show continuous eclipses and typically have shorter periods than detached EBs.}
    \label{fig:eb_lcs}
\end{figure*}

The EXBA masks were designed and observed to estimate the unbiased occurrence fraction of EBs in the Kepler field. 
Although in this work we do not pursue a thorough and complete search of this type of variables, during the exoplanet search previously describe we found many EBs. 
Here we present a catalog of EBs and their eclipsing parameters derived from the BLS periodogram.
Table \ref{tab:eb_table} lists 69 EB systems found in the EXBA masks. 
The model parameters period, transit depth, reference time, and duration are derived from the best BLS fitting.
We perform a visual inspection to remove clear contaminated sources and a cross-reference search with the literature (Simbad search).
The two types of EBs are present in this catalog. 
One of them are contact systems (EWs, see Figure \ref{fig:eb_lcs}) where both components are in contact and sharing a common envelope. 
EWs have light curves in which both minima are of similar depth and their typical periods are shorter than a day showing large-amplitude variability. 
The second type are detached systems (EAs, see Figure \ref{fig:eb_lcs}), where both components are of different spectral type.
Therefore, both the primary and secondary minima have different depths, and the beginning and end of the eclipses are typically very well defined. 
EAs are the major source of contamination when searching for exoplanets using the transiting method.

\clearpage
\startlongtable
\begin{deluxetable*}{lrrrr}
\tablecaption{Catalog of 69 Eclipsing Binary systems found in the EXBA masks. This EB catalog represents an incomplete search for this type of variables and is only presented as an illustration of the opportunities that this data presents. These EBs are a result of a search for periodic transiting signal limited to periods between 0.2 and 100 days and with a standardized periodogram peak value larger than 20, see Section \ref{subsubsec:bls} for further details.
\label{tab:eb_table}}
\tablewidth{0pt}
\tablehead{
\colhead{Gaia ID} & \colhead{Period} & \colhead{$BKJD_0$} & \colhead{Duration} & \colhead{Depth} \\
\colhead{} & \colhead{days} & \colhead{days} &\colhead{days} &\colhead{$\%$}
}
\startdata
    Gaia EDR3 2051774184464360064 &    0.816420 &        443.893508 &         0.099 &   1.030747 \\
    Gaia EDR3 2051871182004752512 &    0.531408 &        443.762933 &         0.450 &   2.848130 \\
    Gaia EDR3 2052499488484831616 &    0.523529 &        630.445491 &         0.450 &   6.528912 \\
    Gaia EDR3 2052503164977145088 &    0.551766 &        443.795923 &         0.450 &   0.104680 \\
    Gaia EDR3 2052503164977145472 &    0.551766 &        443.795923 &         0.450 &  12.661261 \\
    Gaia EDR3 2052524429346526464 &    0.969631 &        540.414513 &         0.099 &   9.448619 \\
    Gaia EDR3 2053614118393506176 &    0.532076 &        443.725578 &         0.099 &  18.842998 \\
    Gaia EDR3 2053614182812468480 &    0.532076 &        443.725578 &         0.099 &   0.997837 \\
    Gaia EDR3 2073608428123884416 &    0.707867 &        443.657609 &         0.150 &   0.096724 \\
    Gaia EDR3 2073608462478158848 &    2.006032 &        445.105109 &         0.099 &   0.198396 \\
    Gaia EDR3 2073608462483635968 &    2.006032 &        445.105109 &         0.099 &   0.205073 \\
    Gaia EDR3 2073682438989499392 &    0.707867 &        443.660681 &         0.126 &   0.056299 \\
    Gaia EDR3 2073682438989966592 &    0.707867 &        443.660681 &         0.126 &   0.042618 \\
    Gaia EDR3 2073682640842559360 &    0.707867 &        630.538708 &         0.126 &   5.968305 \\
    Gaia EDR3 2073682645147919616 &    0.707867 &        443.840681 &         0.450 &   7.558699 \\
    Gaia EDR3 2075404789599593088 &    3.034506 &        444.869691 &         0.330 &   2.204107 \\
    Gaia EDR3 2075405133185880704 &    0.541367 &        443.837691 &         0.450 &   0.821595 \\
    Gaia EDR3 2075405133185894656 &    0.541367 &        443.837691 &         0.450 &   0.230565 \\
    Gaia EDR3 2075406541944219520 &    0.630683 &        443.765691 &         0.450 &   0.072754 \\
    Gaia EDR3 2076208292075451136 &    0.646481 &        443.890268 &         0.399 &   0.829846 \\
    Gaia EDR3 2076209357217808384 &    0.618034 &        443.722268 &         0.399 &   5.536100 \\
    Gaia EDR3 2076209456001789056 &    6.810999 &        445.073768 &         0.150 &   1.027946 \\
    Gaia EDR3 2076505018468886784 &    0.542118 &        443.714849 &         0.450 &   0.707253 \\
    Gaia EDR3 2076789723264283008 &   13.362883 &        446.483764 &         0.450 &   1.503108 \\
    Gaia EDR3 2076789727566520192 &    3.524663 &        445.685764 &         0.450 &   0.800765 \\
    Gaia EDR3 2076836727387736576 &    3.554231 &        542.873966 &         0.201 &   2.813851 \\
    Gaia EDR3 2076836864825475456 &    0.707867 &        443.663764 &         0.126 &   1.174416 \\
    Gaia EDR3 2077245814431169792 &    0.637804 &        630.807001 &         0.051 &  32.182456 \\
    Gaia EDR3 2077246024892437120 &   12.062975 &        445.235872 &         0.450 &   0.831434 \\
    Gaia EDR3 2077810555389602176 &    0.604800 &        539.753960 &         0.399 &   2.754031 \\
    Gaia EDR3 2078117181696007296 &    0.983511 &        444.397464 &         0.099 &  11.280642 \\
    Gaia EDR3 2078734140854317440 &    0.525966 &        735.642432 &         0.450 &   6.108658 \\
    Gaia EDR3 2078735858841858816 &    2.414770 &        445.490808 &         0.174 &   1.672115 \\
    Gaia EDR3 2079019292324040704 &    0.628500 &        443.743214 &         0.051 &   1.849365 \\
    Gaia EDR3 2079399006085883008 &   11.236515 &        446.155307 &         0.099 &   1.319522 \\
    Gaia EDR3 2080297753770786944 &    2.170456 &        443.921954 &         0.126 &   1.089285 \\
    Gaia EDR3 2080297753770788864 &    0.827853 &        444.160454 &         0.051 &   4.862135 \\
    Gaia EDR3 2082091568328182528 &    0.701906 &        443.740152 &         0.051 &   2.486457 \\
    Gaia EDR3 2086516449796423936 &    2.632309 &        443.777988 &         0.174 &  18.155153 \\
    Gaia EDR3 2086516454099067776 &    2.632309 &        443.780988 &         0.126 &   0.191745 \\
    Gaia EDR3 2086592354758573568 &    6.359473 &        444.454549 &         0.399 &   1.276964 \\
    Gaia EDR3 2100240347774313728 &    1.672736 &        443.540000 &         0.099 &  10.979652 \\
    Gaia EDR3 2100490555389337984 &    0.580361 &        443.824932 &         0.051 &  11.126328 \\
    Gaia EDR3 2101033920292227584 &    3.734361 &        447.016742 &         0.051 &   6.638635 \\
    Gaia EDR3 2101136415390982528 &    0.752610 &        443.745168 &         0.450 &   1.506718 \\
    Gaia EDR3 2101136419694167040 &    0.752610 &        443.754168 &         0.450 &   1.984499 \\
    Gaia EDR3 2101752936479788544 &    0.552366 &        443.755662 &         0.051 &   9.413652 \\
    Gaia EDR3 2101753516292218624 &    0.530687 &        443.817162 &         0.450 &   7.978087 \\
    Gaia EDR3 2101753520591885568 &    0.530687 &        443.817162 &         0.450 &   7.094880 \\
    Gaia EDR3 2101753619371695872 &    0.885682 &        444.205662 &         0.099 &   4.122006 \\
    Gaia EDR3 2102900959754861440 &    1.481648 &        443.734874 &         0.099 &   3.322106 \\
    Gaia EDR3 2103827298302304128 &    0.533752 &        539.746877 &         0.450 &   1.101425 \\
    Gaia EDR3 2104814389164263168 &    0.595171 &        443.855354 &         0.051 &   4.316045 \\
    Gaia EDR3 2106327622103110784 &   20.544483 &        463.879921 &         0.450 &   5.975260 \\
    Gaia EDR3 2107265956496766208 &    0.604584 &        443.898714 &         0.450 &   5.087825 \\
    Gaia EDR3 2107266536313762816 &    1.640355 &        444.341214 &         0.099 &  12.684281 \\
    Gaia EDR3 2126083788772171648 &    0.608346 &        539.695481 &         0.450 &   0.021125 \\
    Gaia EDR3 2126085369316699648 &    0.601366 &        539.876981 &         0.399 &   2.233859 \\
    Gaia EDR3 2126085369320144640 &    0.841974 &        539.699981 &         0.051 &   3.655029 \\
    Gaia EDR3 2126955323535081728 &    0.608346 &        539.695419 &         0.450 &  16.312343 \\
    Gaia EDR3 2127956149696086272 &    0.601936 &        539.759749 &         0.399 &   1.934922 \\
    Gaia EDR3 2127957734541969152 &    0.658113 &        443.884613 &         0.051 &  10.867783 \\
    Gaia EDR3 2128160285200885888 &    0.562013 &        443.844143 &         0.450 &   5.224848 \\
    Gaia EDR3 2128267968619535872 &    0.515723 &        443.959706 &         0.051 &   0.231262 \\
    Gaia EDR3 2128268587094813056 &    0.585238 &        443.715206 &         0.450 &   1.411582 \\
    Gaia EDR3 2129514093249096832 &    0.875163 &        444.168417 &         0.450 &   1.554462 \\
    Gaia EDR3 2130912843839236224 &   25.235131 &        460.013006 &         0.249 &   3.355718 \\
    Gaia EDR3 2133602451140200192 &    0.985036 &        444.216754 &         0.174 &   2.357031 \\
    Gaia EDR3 2134814804448660608 &    0.707178 &        444.006163 &         0.150 &  10.898769 \\
\enddata
\tablecomments{The $BKJD_0$, duration, and depth are the results from the best BLS model and account for the primary eclipse. The $BKJD_0$ correspond to the Barycentric Kepler Julian Day ($BKJD = BJD -2,454,833.0$). These parameters were obtained from fitting the BLS model with a broad period and duration grid, therefore are not necessarily the most accurate results.}
\end{deluxetable*}

\section{Discussion} \label{sec:discurssion}

In this section, we will discuss the relevant assumptions of our work as well as its limitations.

The use of custom apertures that follow the PRF shape becomes important when its profile are elongated and a significant fraction of the flux falls in the wings of the profile.
Elongated PRF profiles are particularly true for \textit{Kepler} as we move away from the center of the focal plane (see Figure \ref{fig:prf_ffi}).
Although these aperture masks are more complex than standard circular-like apertures, photometry in crowded regions where multiple sources are located within pixel distance is sub-optimal. 
Under these circumstances, PSF/PRF photometry is recommended, such as the LFD photometry method.
Nonetheless, for isolated sources, custom apertures represent a fast and accurate method to extract their photometry.

As discussed in \cite{2021AJ....162..107H}, fitting the PRF model follows a series of assumptions.
Two of them are relevant to our work: the PRF does not change with time during each quarter and is uniform across each channel.
Both assumptions are reasonable up to some limit, the PRF shape can suffer changes in time due to focus recalibration of the instrument and spatially within a channel. 

The time dependency can be mostly corrected using the CBVs as they capture the instrument systematics.
Another approach is to introduce a time dependency to the PRF model, although this can not be done using \textit{Kepler}'s FFIs due to their limited observing cadence. 
This strategy is feasible if using \textit{Kepler}'s clusters apertures (e.g. NGC 6819), where the number of sources is high enough to provide a smooth PRF model ($\sim 2,000$ sources and $\sim 40,000$ pixel data points) and the observed cadence is in long mode ($\sim 4400$ cadences).

The PRF profile can slightly change across a single channel, these variations are averaged when estimating the shape model for the full CCD. As the typical scale of a PRF model is $< 4$ pixels and its variations within the channel happen at a smaller scale (1 to 2 pixels), the effect in the photometric apertures defined from the PRF profile is minimal, especially for the small size of the EXBA masks (36 x 60 pixels) compared to the FFIs ($1,024$ x $1,100$ pixels).
If needed, the spatial change of the PRF profile can be model by including a spatial dependency to the PRF model or by dividing the channel image into a grid where each cell contains a dedicated model of the PRF.

The catalog of light curves for $9,327$ sources presented in this work is based on a Gaia search done at a limiting magnitude of $G = 20$. 
Thus, fainter sources not only do not have time series but also are not considered for contamination metrics. 
This sets the CROWDSAP metric reported in this work as an upper limit, but the flux contribution from sources fainter than 20th magnitude is minimal in \textit{Kepler} observations.
The Gaia search could be relaxed in future works to allow fainter sources.

\section{Summary} \label{sec:summary}

In this work, we extracted source light curves from \textit{Kepler}'s EXBA masks for the first time. 
The EXBA masks are a set of custom masks observed by \textit{Kepler} during its prime mission between quarters 5 and 17, which has never been analyzed before. 
We created light curve files for $9,327$ Gaia sources observed in the EXBA masks. 
In our processing pipeline, we used the method described in the LFD photometry \citep{2021AJ....162..107H} to model the PRF profile of all channels using \textit{Kepler}'s full frame images.
We capitalized on the large number of sources observed in these FFIs ($\sim 10,000$ per channel) to obtained detailed PRF models.
We evaluated the PRF models for every source in the EXBA masks to define photometric aperture. 
These aperture masks follow the characteristics shapes of the PRF and enabled to compute flux completeness and crowdedness metrics. 
The light curves files produced in this work are available for public access through the MAST archive. 
The PRF profile models built from the FFIs are a sub-product of our work. 
The \texttt{Python} library \texttt{kepler-apertures} enables user to utilize our precomputed PRF models (included in the package) to perform aperture photometry on \textit{Kepler} like data.

As an example of what can be accomplished with these high-level science products, we present two scientific cases. 
First, we find evidence of an exoplanet candidate around \candidate. 
The results of the transit model fitting suggest that the object producing this transit has a size smaller than $2.2 R_J$.
Further investigations need to be conducted to constrain the stellar properties of the host as well as the parameters of the transiting objects in order to confirm its planetary nature.
As a second case, we presented a catalog of Eclipsing Binaries found during the search for transiting signals. 
The EXBA masks were designed with the primary purpose of estimating the fraction of background EBs systems in the \textit{Kepler} field.
EBs are the major source of false positives when searching for exoplanets. 
Although in this work we did not attempt a complete search of EBs, we found a significant number (69) of them and compute its basic parameters using the BLS periodogram. 
A more thorough search and analysis of the EXBA light curves to estimate the occurrence fraction of EBs is left to the community.

This work presents the tools to perform aperture photometry of \textit{Kepler} data in a similar fashion as the \textit{Kepler} pipeline and provides to the community a new dataset that has not been systematically explored yet. 
This methodology can be applied systematically to all the background sources found in the \textit{Kepler} targets, which are estimated to contain more than $129,000$ sources.
Although a large fraction of them are blended sources and a fully PSF photometry approach is more suitable (e.g. LFD photometry), numerous of these sources are isolated, and fast aperture photometry is the optimal approach.
Even more, the \textit{K2} mission shares similar data characteristics as \textit{Kepler}'s prime mission and the application of these tools extends naturally.

\begin{acknowledgments}

This paper includes data collected by the Kepler mission and obtained from the MAST data archive at the Space Telescope Science Institute (STScI). Funding for the Kepler mission is provided by the NASA Science Mission Directorate. STScI is operated by the Association of Universities for Research in Astronomy, Inc., under NASA contract NAS 5–26555. This work has made use of data from the European Space Agency (ESA) mission Gaia (\url{https://www. cosmos.esa.int/gaia}), processed by the Gaia Data Processing and Analysis Consortium (DPAC, \url{https://www.cosmos.esa. int/web/gaia/dpac/consortium}). Funding for the DPAC has been provided by national institutions, in particular the institutions participating in the Gaia Multilateral Agreement. This research made use of exoplanet \citep{exoplanet:exoplanet} and its dependencies\citep{2013A&A...558A..33A, 2018AJ....156..123A, 2013MNRAS.435.2152K, 2016arXiv160502688T, 10.7717/peerj-cs.55, 2019AJ....157...64L, Kumar2019, Agol_2020, exoplanet:exoplanet}. Funding for this work for JMP and CH is provided by grant number 80NSSC20K0874, through NASA ROSES.

\end{acknowledgments}

%

\vspace{5mm}
\facilities{Kepler}


\software{astropy \citep{2013A&A...558A..33A},  
          lightkurve \citep{2018ascl.soft12013L}, 
          scipy \citep{2020SciPy-NMeth},
          psfmachine \citep{2021zndo...4784073H},
          kepler-apertures \citep{jorge_martinez_palomera_2021_5062871},
          numpy \citep{harris2020array},
          exoplanet \citep{exoplanet:exoplanet},
          pymc3 \citep{10.7717/peerj-cs.55},
          contaminante \citep{2021RNAAS...5..260H},
          }



\appendix

\section{EXBA Light Curve Files} \label{appx:lcfs}

Each EXBA Light Curve File (LCF) contains the light curve of a single source and the aperture masks used for the photometry of a single quarter. Each LCF is named under the following pattern \texttt{hlsp\_kbonus-apexba\_kepler\_kepler\_<gaia\_source\_id>-q<quarter>\_kepler\_v1.0\_lc.fits} where the fields are the Gaia EDR3 \textit{source id} and the quarter number, respectively. Each LCF is a multi-extension FITS file that contains the following:

\begin{itemize}
    \item The primary header unit. This extension contain the main metadata necessary to identify the object.
    \item The LIGHTCURVE extension. A binary table header unit containing the flux time series information. The columns of the table are detailed in Table \ref{tab:lc_tab}.
    \item The APERTURE extensions. Eight image header units containing the aperture masks used for the eight different apertures as described in Section \ref{subsec:aper}: first an optimized aperture (0) and then 7 apertures with decreasing size as percentile increase. Each aperture mask matches the shape of the 4 tiled EXBA masks (36 x 60 pixels) for a given channel.
\end{itemize}

Accompanying the EXBA LCFs for a given quarter and channel combination, there are the \texttt{hlsp\_kbonus-apexba\_kepler\_kepler\_exba-mask-ch<channel>-q<quarter>\_kepler\_v1.0\_sup.fits} files. These are support files used for plotting purposes. Each file contains three extensions:

\begin{itemize}
    \item The primary header unit containing basic information of the EXBA mask such as center R.A. and Dec coordinates and pixel reference position.
    \item An image header unit with an averaged image across cadences of the tiled EXBA mask (36 x 60 pixels).
    \item A binary table unit containing the Gaia EDR3 catalog of sources observed in the EXBA mask.
\end{itemize}

\begin{deluxetable*}{lllcl}[htb!]
\tablecaption{Description of the columns available in the Binary Table header unit containing the flux time series of sources in every EXBA Light Curve File.
\label{tab:lc_tab}}
\tablewidth{0pt}
\tablehead{
\colhead{Column} & \colhead{Field} &\colhead{Format} &\colhead{Units} &\colhead{Description}
}
\startdata
 1 & Time          & float32  & BJD - 2454833 & Time value. \\
 2 & Cadenceno     & int64    & -      & Cadence number. \\
 3 & Flux          & float32  & $e^-/s$ & Flux from the optimized aperture. \\
 4 & Flux0         & float32  & $e^-/s$ & Flux from the 1st aperture at percentile 0. \\
 5 & Flux15        & float32  & $e^-/s$ & Flux from the 2nd aperture at percentile 15. \\
 6 & Flux30        & float32  & $e^-/s$ & Flux from the 3rd aperture at percentile 30. \\
 7 & Flux45        & float32  & $e^-/s$ & Flux from the 4th aperture at percentile 45. \\
 8 & Flux60        & float32  & $e^-/s$ & Flux from the 5th aperture at percentile 60. \\
 9 & Flux75        & float32  & $e^-/s$ & Flux from the 6th aperture at percentile 75. \\
10 & Flux90        & float32  & $e^-/s$ & Flux from the 7th aperture at percentile 90. \\
11 & Flux\_err     & float32  & $e^-/s$ & Flux error from the optimized aperture. \\
12 & Flux\_err0    & float32  & $e^-/s$ & Flux error from the 1st aperture at percentile 0. \\
13 & Flux\_err15   & float32  & $e^-/s$ & Flux error from the 2nd aperture at percentile 15. \\
14 & Flux\_err30   & float32  & $e^-/s$ & Flux error from the 3rd aperture at percentile 30. \\
15 & Flux\_err45   & float32  & $e^-/s$ & Flux error from the 4th aperture at percentile 45. \\
16 & Flux\_err60   & float32  & $e^-/s$ & Flux error from the 5th aperture at percentile 60. \\
17 & Flux\_err75   & float32  & $e^-/s$ & Flux error from the 6th aperture at percentile 75. \\
18 & Flux\_err90   & float32  & $e^-/s$ & Flux error from the 7th aperture at percentile 90. \\
19 & Quality       & int64    & -      & Quality flag from the TPF. \\
20 & SAP\_Quality  & int64    & -      & Quality flag introduced by \texttt{lightkurve.KeplerLightCurve}. \\
\enddata
\end{deluxetable*}
    


\bibliography{bibtex.bib}

\begin{thebibliography}{}
\expandafter\ifx\csname natexlab\endcsname\relax\def\natexlab#1{#1}\fi
\providecommand{\url}[1]{\href{#1}{#1}}
\providecommand{\dodoi}[1]{doi:~\href{http://doi.org/#1}{\nolinkurl{#1}}}
\providecommand{\doeprint}[1]{\href{http://ascl.net/#1}{\nolinkurl{http://ascl.net/#1}}}
\providecommand{\doarXiv}[1]{\href{https://arxiv.org/abs/#1}{\nolinkurl{https://arxiv.org/abs/#1}}}

\bibitem[{Agol {et~al.}(2020)Agol, Luger, \& Foreman-Mackey}]{Agol_2020}
Agol, E., Luger, R., \& Foreman-Mackey, D. 2020, The Astronomical Journal, 159,
  123, \dodoi{10.3847/1538-3881/ab4fee}

\bibitem[{{Astropy Collaboration} {et~al.}(2013){Astropy Collaboration},
  {Robitaille}, {Tollerud}, {Greenfield}, {Droettboom}, {Bray}, {Aldcroft},
  {Davis}, {Ginsburg}, {Price-Whelan}, {Kerzendorf}, {Conley}, {Crighton},
  {Barbary}, {Muna}, {Ferguson}, {Grollier}, {Parikh}, {Nair}, {Unther},
  {Deil}, {Woillez}, {Conseil}, {Kramer}, {Turner}, {Singer}, {Fox}, {Weaver},
  {Zabalza}, {Edwards}, {Azalee Bostroem}, {Burke}, {Casey}, {Crawford},
  {Dencheva}, {Ely}, {Jenness}, {Labrie}, {Lim}, {Pierfederici}, {Pontzen},
  {Ptak}, {Refsdal}, {Servillat}, \& {Streicher}}]{2013A&A...558A..33A}
{Astropy Collaboration}, {Robitaille}, T.~P., {Tollerud}, E.~J., {et~al.} 2013,
  \aap, 558, A33, \dodoi{10.1051/0004-6361/201322068}

\bibitem[{{Astropy Collaboration} {et~al.}(2018){Astropy Collaboration},
  {Price-Whelan}, {Sip{\H{o}}cz}, {G{\"u}nther}, {Lim}, {Crawford}, {Conseil},
  {Shupe}, {Craig}, {Dencheva}, {Ginsburg}, {VanderPlas}, {Bradley},
  {P{\'e}rez-Su{\'a}rez}, {de Val-Borro}, {Aldcroft}, {Cruz}, {Robitaille},
  {Tollerud}, {Ardelean}, {Babej}, {Bach}, {Bachetti}, {Bakanov}, {Bamford},
  {Barentsen}, {Barmby}, {Baumbach}, {Berry}, {Biscani}, {Boquien}, {Bostroem},
  {Bouma}, {Brammer}, {Bray}, {Breytenbach}, {Buddelmeijer}, {Burke},
  {Calderone}, {Cano Rodr{\'\i}guez}, {Cara}, {Cardoso}, {Cheedella}, {Copin},
  {Corrales}, {Crichton}, {D'Avella}, {Deil}, {Depagne}, {Dietrich}, {Donath},
  {Droettboom}, {Earl}, {Erben}, {Fabbro}, {Ferreira}, {Finethy}, {Fox},
  {Garrison}, {Gibbons}, {Goldstein}, {Gommers}, {Greco}, {Greenfield},
  {Groener}, {Grollier}, {Hagen}, {Hirst}, {Homeier}, {Horton}, {Hosseinzadeh},
  {Hu}, {Hunkeler}, {Ivezi{\'c}}, {Jain}, {Jenness}, {Kanarek}, {Kendrew},
  {Kern}, {Kerzendorf}, {Khvalko}, {King}, {Kirkby}, {Kulkarni}, {Kumar},
  {Lee}, {Lenz}, {Littlefair}, {Ma}, {Macleod}, {Mastropietro}, {McCully},
  {Montagnac}, {Morris}, {Mueller}, {Mumford}, {Muna}, {Murphy}, {Nelson},
  {Nguyen}, {Ninan}, {N{\"o}the}, {Ogaz}, {Oh}, {Parejko}, {Parley}, {Pascual},
  {Patil}, {Patil}, {Plunkett}, {Prochaska}, {Rastogi}, {Reddy Janga},
  {Sabater}, {Sakurikar}, {Seifert}, {Sherbert}, {Sherwood-Taylor}, {Shih},
  {Sick}, {Silbiger}, {Singanamalla}, {Singer}, {Sladen}, {Sooley},
  {Sornarajah}, {Streicher}, {Teuben}, {Thomas}, {Tremblay}, {Turner},
  {Terr{\'o}n}, {van Kerkwijk}, {de la Vega}, {Watkins}, {Weaver}, {Whitmore},
  {Woillez}, {Zabalza}, \& {Astropy Contributors}}]{2018AJ....156..123A}
{Astropy Collaboration}, {Price-Whelan}, A.~M., {Sip{\H{o}}cz}, B.~M., {et~al.}
  2018, \aj, 156, 123, \dodoi{10.3847/1538-3881/aabc4f}

\bibitem[{{Bailer-Jones} {et~al.}(2021){Bailer-Jones}, {Rybizki}, {Fouesneau},
  {Demleitner}, \& {Andrae}}]{2021AJ....161..147B}
{Bailer-Jones}, C.~A.~L., {Rybizki}, J., {Fouesneau}, M., {Demleitner}, M., \&
  {Andrae}, R. 2021, \aj, 161, 147, \dodoi{10.3847/1538-3881/abd806}

\bibitem[{{Balona} {et~al.}(2013){Balona}, {Medupe}, {Abedigamba}, {Ayane},
  {Keeley}, {Matsididi}, {Mekonnen}, {Nhlapo}, \&
  {Sithole}}]{2013MNRAS.430.3472B}
{Balona}, L.~A., {Medupe}, T., {Abedigamba}, O.~P., {et~al.} 2013, \mnras, 430,
  3472, \dodoi{10.1093/mnras/stt148}

\bibitem[{{Barentsen} {et~al.}(2018){Barentsen}, {Hedges}, {Saunders}, {Cody},
  {Gully-Santiago}, {Bryson}, \& {Dotson}}]{2018arXiv181012554B}
{Barentsen}, G., {Hedges}, C., {Saunders}, N., {et~al.} 2018, arXiv e-prints,
  arXiv:1810.12554.
\newblock \doarXiv{1810.12554}

\bibitem[{{Batalha} {et~al.}(2010){Batalha}, {Borucki}, {Koch}, {Bryson},
  {Haas}, {Brown}, {Caldwell}, {Hall}, {Gilliland}, {Latham}, {Meibom}, \&
  {Monet}}]{2010ApJ...713L.109B}
{Batalha}, N.~M., {Borucki}, W.~J., {Koch}, D.~G., {et~al.} 2010, \apjl, 713,
  L109, \dodoi{10.1088/2041-8205/713/2/L109}

\bibitem[{Belokurov {et~al.}(2020)Belokurov, Penoyre, Oh, Iorio, Hodgkin,
  Evans, Everall, Koposov, Tout, Izzard, Clarke, \&
  Brown}]{10.1093/mnras/staa1522}
Belokurov, V., Penoyre, Z., Oh, S., {et~al.} 2020, Monthly Notices of the Royal
  Astronomical Society, 496, 1922, \dodoi{10.1093/mnras/staa1522}

\bibitem[{{Borucki} {et~al.}(2010){Borucki}, {Koch}, {Basri}, {Batalha},
  {Brown}, {Caldwell}, {Caldwell}, {Christensen-Dalsgaard}, {Cochran},
  {DeVore}, {Dunham}, {Dupree}, {Gautier}, {Geary}, {Gilliland}, {Gould},
  {Howell}, {Jenkins}, {Kondo}, {Latham}, {Marcy}, {Meibom}, {Kjeldsen},
  {Lissauer}, {Monet}, {Morrison}, {Sasselov}, {Tarter}, {Boss}, {Brownlee},
  {Owen}, {Buzasi}, {Charbonneau}, {Doyle}, {Fortney}, {Ford}, {Holman},
  {Seager}, {Steffen}, {Welsh}, {Rowe}, {Anderson}, {Buchhave}, {Ciardi},
  {Walkowicz}, {Sherry}, {Horch}, {Isaacson}, {Everett}, {Fischer}, {Torres},
  {Johnson}, {Endl}, {MacQueen}, {Bryson}, {Dotson}, {Haas}, {Kolodziejczak},
  {Van Cleve}, {Chandrasekaran}, {Twicken}, {Quintana}, {Clarke}, {Allen},
  {Li}, {Wu}, {Tenenbaum}, {Verner}, {Bruhweiler}, {Barnes}, \&
  {Prsa}}]{2010Sci...327..977B}
{Borucki}, W.~J., {Koch}, D., {Basri}, G., {et~al.} 2010, Science, 327, 977,
  \dodoi{10.1126/science.1185402}

\bibitem[{{Brewer} {et~al.}(2013){Brewer}, {Sandquist}, {Mathieu}, {Milliman},
  {Geller}, {Jeffries}, {Orosz}, {Brogaard}, {Platais}, {Bruntt}, {Grundahl},
  {Stello}, \& {Frandsen}}]{2013AAS...22125036B}
{Brewer}, L., {Sandquist}, E.~L., {Mathieu}, R.~D., {et~al.} 2013, in American
  Astronomical Society Meeting Abstracts, Vol. 221, American Astronomical
  Society Meeting Abstracts \#221, 250.36

\bibitem[{{Brown} {et~al.}(2011){Brown}, {Latham}, {Everett}, \&
  {Esquerdo}}]{2011AJ....142..112B}
{Brown}, T.~M., {Latham}, D.~W., {Everett}, M.~E., \& {Esquerdo}, G.~A. 2011,
  \aj, 142, 112, \dodoi{10.1088/0004-6256/142/4/112}

\bibitem[{{Bryson} {et~al.}(2010){Bryson}, {Tenenbaum}, {Jenkins},
  {Chandrasekaran}, {Klaus}, {Caldwell}, {Gilliland}, {Haas}, {Dotson}, {Koch},
  \& {Borucki}}]{2010ApJ...713L..97B}
{Bryson}, S.~T., {Tenenbaum}, P., {Jenkins}, J.~M., {et~al.} 2010, \apjl, 713,
  L97, \dodoi{10.1088/2041-8205/713/2/L97}

\bibitem[{{Choi} {et~al.}(2016){Choi}, {Dotter}, {Conroy}, {Cantiello},
  {Paxton}, \& {Johnson}}]{2016ApJ...823..102C}
{Choi}, J., {Dotter}, A., {Conroy}, C., {et~al.} 2016, \apj, 823, 102,
  \dodoi{10.3847/0004-637X/823/2/102}

\bibitem[{{Cody} {et~al.}(2018){Cody}, {Barentsen}, {Hedges}, {Gully-Santiago},
  \& {Cardoso}}]{2018RNAAS...2Q..25C}
{Cody}, A.~M., {Barentsen}, G., {Hedges}, C., {Gully-Santiago}, M., \&
  {Cardoso}, J. V. d.~M. 2018, Research Notes of the American Astronomical
  Society, 2, 25, \dodoi{10.3847/2515-5172/aaac30}

\bibitem[{{Corsaro} {et~al.}(2012){Corsaro}, {Stello}, {Huber}, {Bedding},
  {Bonanno}, {Brogaard}, {Kallinger}, {Benomar}, {White}, {Mosser}, {Basu},
  {Chaplin}, {Christensen-Dalsgaard}, {Elsworth}, {Garc{\'\i}a}, {Hekker},
  {Kjeldsen}, {Mathur}, {Meibom}, {Hall}, {Ibrahim}, \&
  {Klaus}}]{2012ApJ...757..190C}
{Corsaro}, E., {Stello}, D., {Huber}, D., {et~al.} 2012, \apj, 757, 190,
  \dodoi{10.1088/0004-637X/757/2/190}

\bibitem[{{Cutri} \& {et al.}(2012)}]{2012yCat.2311....0C}
{Cutri}, R.~M., \& {et al.} 2012, VizieR Online Data Catalog, II/311

\bibitem[{{Cutri} {et~al.}(2003){Cutri}, {Skrutskie}, {van Dyk}, {Beichman},
  {Carpenter}, {Chester}, {Cambresy}, {Evans}, {Fowler}, {Gizis}, {Howard},
  {Huchra}, {Jarrett}, {Kopan}, {Kirkpatrick}, {Light}, {Marsh}, {McCallon},
  {Schneider}, {Stiening}, {Sykes}, {Weinberg}, {Wheaton}, {Wheelock}, \&
  {Zacarias}}]{2003yCat.2246....0C}
{Cutri}, R.~M., {Skrutskie}, M.~F., {van Dyk}, S., {et~al.} 2003, VizieR Online
  Data Catalog, II/246

\bibitem[{{Dotter}(2016)}]{2016ApJS..222....8D}
{Dotter}, A. 2016, \apjs, 222, 8, \dodoi{10.3847/0067-0049/222/1/8}

\bibitem[{{Eastman} {et~al.}(2013){Eastman}, {Gaudi}, \&
  {Agol}}]{2013PASP..125...83E}
{Eastman}, J., {Gaudi}, B.~S., \& {Agol}, E. 2013, \pasp, 125, 83,
  \dodoi{10.1086/669497}

\bibitem[{{Eastman} {et~al.}(2019){Eastman}, {Rodriguez}, {Agol}, {Stassun},
  {Beatty}, {Vanderburg}, {Gaudi}, {Collins}, \& {Luger}}]{2019arXiv190709480E}
{Eastman}, J.~D., {Rodriguez}, J.~E., {Agol}, E., {et~al.} 2019, arXiv
  e-prints, arXiv:1907.09480.
\newblock \doarXiv{1907.09480}

\bibitem[{{Espinoza} {et~al.}(2019){Espinoza}, {Kossakowski}, \&
  {Brahm}}]{2019MNRAS.490.2262E}
{Espinoza}, N., {Kossakowski}, D., \& {Brahm}, R. 2019, \mnras, 490, 2262,
  \dodoi{10.1093/mnras/stz2688}

\bibitem[{{Fabricius} {et~al.}(2021){Fabricius}, {Luri}, {Arenou}, {Babusiaux},
  {Helmi}, {Muraveva}, {Reyl{\'e}}, {Spoto}, {Vallenari}, {Antoja}, {Balbinot},
  {Barache}, {Bauchet}, {Bragaglia}, {Busonero}, {Cantat-Gaudin}, {Carrasco},
  {Diakit{\'e}}, {Fabrizio}, {Figueras}, {Garcia-Gutierrez}, {Garofalo},
  {Jordi}, {Kervella}, {Khanna}, {Leclerc}, {Licata}, {Lambert}, {Marrese},
  {Masip}, {Ramos}, {Robichon}, {Robin}, {Romero-G{\'o}mez}, {Rubele}, \&
  {Weiler}}]{2021A&A...649A...5F}
{Fabricius}, C., {Luri}, X., {Arenou}, F., {et~al.} 2021, \aap, 649, A5,
  \dodoi{10.1051/0004-6361/202039834}

\bibitem[{Foreman-Mackey {et~al.}(2021)Foreman-Mackey, Savel, Luger, Czekala,
  Agol, Price-Whelan, Hedges, Gilbert, Barclay, Bouma, \&
  Brandt}]{exoplanet:exoplanet}
Foreman-Mackey, D., Savel, A., Luger, R., {et~al.} 2021,
  exoplanet-dev/exoplanet v0.4.5, \dodoi{10.5281/zenodo.1998447}

\bibitem[{{Gaia Collaboration} {et~al.}(2018){Gaia Collaboration}, {Brown},
  {Vallenari}, {Prusti}, {de Bruijne}, {Babusiaux}, {Bailer-Jones}, {Biermann},
  {Evans}, {Eyer}, {Jansen}, {Jordi}, {Klioner}, {Lammers}, {Lindegren},
  {Luri}, {Mignard}, {Panem}, {Pourbaix}, {Randich}, {Sartoretti}, {Siddiqui},
  {Soubiran}, {van Leeuwen}, {Walton}, {Arenou}, {Bastian}, {Cropper},
  {Drimmel}, {Katz}, {Lattanzi}, {Bakker}, {Cacciari}, {Casta{\~n}eda},
  {Chaoul}, {Cheek}, {De Angeli}, {Fabricius}, {Guerra}, {Holl}, {Masana},
  {Messineo}, {Mowlavi}, {Nienartowicz}, {Panuzzo}, {Portell}, {Riello},
  {Seabroke}, {Tanga}, {Th{\'e}venin}, {Gracia-Abril}, {Comoretto},
  {Garcia-Reinaldos}, {Teyssier}, {Altmann}, {Andrae}, {Audard},
  {Bellas-Velidis}, {Benson}, {Berthier}, {Blomme}, {Burgess}, {Busso},
  {Carry}, {Cellino}, {Clementini}, {Clotet}, {Creevey}, {Davidson}, {De
  Ridder}, {Delchambre}, {Dell'Oro}, {Ducourant},
  {Fern{\'a}ndez-Hern{\'a}ndez}, {Fouesneau}, {Fr{\'e}mat}, {Galluccio},
  {Garc{\'\i}a-Torres}, {Gonz{\'a}lez-N{\'u}{\~n}ez}, {Gonz{\'a}lez-Vidal},
  {Gosset}, {Guy}, {Halbwachs}, {Hambly}, {Harrison}, {Hern{\'a}ndez},
  {Hestroffer}, {Hodgkin}, {Hutton}, {Jasniewicz}, {Jean-Antoine-Piccolo},
  {Jordan}, {Korn}, {Krone-Martins}, {Lanzafame}, {Lebzelter}, {L{\"o}ffler},
  {Manteiga}, {Marrese}, {Mart{\'\i}n-Fleitas}, {Moitinho}, {Mora}, {Muinonen},
  {Osinde}, {Pancino}, {Pauwels}, {Petit}, {Recio-Blanco}, {Richards},
  {Rimoldini}, {Robin}, {Sarro}, {Siopis}, {Smith}, {Sozzetti}, {S{\"u}veges},
  {Torra}, {van Reeven}, {Abbas}, {Abreu Aramburu}, {Accart}, {Aerts},
  {Altavilla}, {{\'A}lvarez}, {Alvarez}, {Alves}, {Anderson}, {Andrei},
  {Anglada Varela}, {Antiche}, {Antoja}, {Arcay}, {Astraatmadja}, {Bach},
  {Baker}, {Balaguer-N{\'u}{\~n}ez}, {Balm}, {Barache}, {Barata}, {Barbato},
  {Barblan}, {Barklem}, {Barrado}, {Barros}, {Barstow}, {Bartholom{\'e}
  Mu{\~n}oz}, {Bassilana}, {Becciani}, {Bellazzini}, {Berihuete}, {Bertone},
  {Bianchi}, {Bienaym{\'e}}, {Blanco-Cuaresma}, {Boch}, {Boeche}, {Bombrun},
  {Borrachero}, {Bossini}, {Bouquillon}, {Bourda}, {Bragaglia}, {Bramante},
  {Breddels}, {Bressan}, {Brouillet}, {Br{\"u}semeister}, {Brugaletta},
  {Bucciarelli}, {Burlacu}, {Busonero}, {Butkevich}, {Buzzi}, {Caffau},
  {Cancelliere}, {Cannizzaro}, {Cantat-Gaudin}, {Carballo}, {Carlucci},
  {Carrasco}, {Casamiquela}, {Castellani}, {Castro-Ginard}, {Charlot},
  {Chemin}, {Chiavassa}, {Cocozza}, {Costigan}, {Cowell}, {Crifo}, {Crosta},
  {Crowley}, {Cuypers}, {Dafonte}, {Damerdji}, {Dapergolas}, {David}, {David},
  {de Laverny}, {De Luise}, {De March}, {de Martino}, {de Souza}, {de Torres},
  {Debosscher}, {del Pozo}, {Delbo}, {Delgado}, {Delgado}, {Di Matteo},
  {Diakite}, {Diener}, {Distefano}, {Dolding}, {Drazinos}, {Dur{\'a}n},
  {Edvardsson}, {Enke}, {Eriksson}, {Esquej}, {Eynard Bontemps}, {Fabre},
  {Fabrizio}, {Faigler}, {Falc{\~a}o}, {Farr{\`a}s Casas}, {Federici},
  {Fedorets}, {Fernique}, {Figueras}, {Filippi}, {Findeisen}, {Fonti},
  {Fraile}, {Fraser}, {Fr{\'e}zouls}, {Gai}, {Galleti}, {Garabato},
  {Garc{\'\i}a-Sedano}, {Garofalo}, {Garralda}, {Gavel}, {Gavras}, {Gerssen},
  {Geyer}, {Giacobbe}, {Gilmore}, {Girona}, {Giuffrida}, {Glass}, {Gomes},
  {Granvik}, {Gueguen}, {Guerrier}, {Guiraud}, {Guti{\'e}rrez-S{\'a}nchez},
  {Haigron}, {Hatzidimitriou}, {Hauser}, {Haywood}, {Heiter}, {Helmi}, {Heu},
  {Hilger}, {Hobbs}, {Hofmann}, {Holland}, {Huckle}, {Hypki}, {Icardi},
  {Jan{\ss}en}, {Jevardat de Fombelle}, {Jonker}, {Juh{\'a}sz}, {Julbe},
  {Karampelas}, {Kewley}, {Klar}, {Kochoska}, {Kohley}, {Kolenberg},
  {Kontizas}, {Kontizas}, {Koposov}, {Kordopatis}, {Kostrzewa-Rutkowska},
  {Koubsky}, {Lambert}, {Lanza}, {Lasne}, {Lavigne}, {Le Fustec}, {Le
  Poncin-Lafitte}, {Lebreton}, {Leccia}, {Leclerc}, {Lecoeur-Taibi},
  {Lenhardt}, {Leroux}, {Liao}, {Licata}, {Lindstr{\o}m}, {Lister}, {Livanou},
  {Lobel}, {L{\'o}pez}, {Managau}, {Mann}, {Mantelet}, {Marchal}, {Marchant},
  {Marconi}, {Marinoni}, {Marschalk{\'o}}, {Marshall}, {Martino}, {Marton},
  {Mary}, {Massari}, {Matijevi{\v{c}}}, {Mazeh}, {McMillan}, {Messina},
  {Michalik}, {Millar}, {Molina}, {Molinaro}, {Moln{\'a}r}, {Montegriffo},
  {Mor}, {Morbidelli}, {Morel}, {Morris}, {Mulone}, {Muraveva}, {Musella},
  {Nelemans}, {Nicastro}, {Noval}, {O'Mullane}, {Ord{\'e}novic},
  {Ord{\'o}{\~n}ez-Blanco}, {Osborne}, {Pagani}, {Pagano}, {Pailler},
  {Palacin}, {Palaversa}, {Panahi}, {Pawlak}, {Piersimoni}, {Pineau}, {Plachy},
  {Plum}, {Poggio}, {Poujoulet}, {Pr{\v{s}}a}, {Pulone}, {Racero}, {Ragaini},
  {Rambaux}, {Ramos-Lerate}, {Regibo}, {Reyl{\'e}}, {Riclet}, {Ripepi}, {Riva},
  {Rivard}, {Rixon}, {Roegiers}, {Roelens}, {Romero-G{\'o}mez}, {Rowell},
  {Royer}, {Ruiz-Dern}, {Sadowski}, {Sagrist{\`a} Sell{\'e}s}, {Sahlmann},
  {Salgado}, {Salguero}, {Sanna}, {Santana-Ros}, {Sarasso}, {Savietto},
  {Schultheis}, {Sciacca}, {Segol}, {Segovia}, {S{\'e}gransan}, {Shih},
  {Siltala}, {Silva}, {Smart}, {Smith}, {Solano}, {Solitro}, {Sordo}, {Soria
  Nieto}, {Souchay}, {Spagna}, {Spoto}, {Stampa}, {Steele},
  {Steidelm{\"u}ller}, {Stephenson}, {Stoev}, {Suess}, {Surdej}, {Szabados},
  {Szegedi-Elek}, {Tapiador}, {Taris}, {Tauran}, {Taylor}, {Teixeira},
  {Terrett}, {Teyssandier}, {Thuillot}, {Titarenko}, {Torra Clotet}, {Turon},
  {Ulla}, {Utrilla}, {Uzzi}, {Vaillant}, {Valentini}, {Valette}, {van Elteren},
  {Van Hemelryck}, {van Leeuwen}, {Vaschetto}, {Vecchiato}, {Veljanoski},
  {Viala}, {Vicente}, {Vogt}, {von Essen}, {Voss}, {Votruba}, {Voutsinas},
  {Walmsley}, {Weiler}, {Wertz}, {Wevers}, {Wyrzykowski}, {Yoldas},
  {{\v{Z}}erjal}, {Ziaeepour}, {Zorec}, {Zschocke}, {Zucker}, {Zurbach}, \&
  {Zwitter}}]{2018A&A...616A...1G}
{Gaia Collaboration}, {Brown}, A.~G.~A., {Vallenari}, A., {et~al.} 2018, \aap,
  616, A1, \dodoi{10.1051/0004-6361/201833051}

\bibitem[{{Gaia Collaboration} {et~al.}(2021){Gaia Collaboration}, {Brown},
  {Vallenari}, {Prusti}, {de Bruijne}, {Babusiaux}, {Biermann}, {Creevey},
  {Evans}, {Eyer}, {Hutton}, {Jansen}, {Jordi}, {Klioner}, {Lammers},
  {Lindegren}, {Luri}, {Mignard}, {Panem}, {Pourbaix}, {Randich}, {Sartoretti},
  {Soubiran}, {Walton}, {Arenou}, {Bailer-Jones}, {Bastian}, {Cropper},
  {Drimmel}, {Katz}, {Lattanzi}, {van Leeuwen}, {Bakker}, {Cacciari},
  {Casta{\~n}eda}, {De Angeli}, {Ducourant}, {Fabricius}, {Fouesneau},
  {Fr{\'e}mat}, {Guerra}, {Guerrier}, {Guiraud}, {Jean-Antoine Piccolo},
  {Masana}, {Messineo}, {Mowlavi}, {Nicolas}, {Nienartowicz}, {Pailler},
  {Panuzzo}, {Riclet}, {Roux}, {Seabroke}, {Sordo}, {Tanga}, {Th{\'e}venin},
  {Gracia-Abril}, {Portell}, {Teyssier}, {Altmann}, {Andrae}, {Bellas-Velidis},
  {Benson}, {Berthier}, {Blomme}, {Brugaletta}, {Burgess}, {Busso}, {Carry},
  {Cellino}, {Cheek}, {Clementini}, {Damerdji}, {Davidson}, {Delchambre},
  {Dell'Oro}, {Fern{\'a}ndez-Hern{\'a}ndez}, {Galluccio}, {Garc{\'\i}a-Lario},
  {Garcia-Reinaldos}, {Gonz{\'a}lez-N{\'u}{\~n}ez}, {Gosset}, {Haigron},
  {Halbwachs}, {Hambly}, {Harrison}, {Hatzidimitriou}, {Heiter},
  {Hern{\'a}ndez}, {Hestroffer}, {Hodgkin}, {Holl}, {Jan{\ss}en}, {Jevardat de
  Fombelle}, {Jordan}, {Krone-Martins}, {Lanzafame}, {L{\"o}ffler}, {Lorca},
  {Manteiga}, {Marchal}, {Marrese}, {Moitinho}, {Mora}, {Muinonen}, {Osborne},
  {Pancino}, {Pauwels}, {Petit}, {Recio-Blanco}, {Richards}, {Riello},
  {Rimoldini}, {Robin}, {Roegiers}, {Rybizki}, {Sarro}, {Siopis}, {Smith},
  {Sozzetti}, {Ulla}, {Utrilla}, {van Leeuwen}, {van Reeven}, {Abbas}, {Abreu
  Aramburu}, {Accart}, {Aerts}, {Aguado}, {Ajaj}, {Altavilla}, {{\'A}lvarez},
  {{\'A}lvarez Cid-Fuentes}, {Alves}, {Anderson}, {Anglada Varela}, {Antoja},
  {Audard}, {Baines}, {Baker}, {Balaguer-N{\'u}{\~n}ez}, {Balbinot}, {Balog},
  {Barache}, {Barbato}, {Barros}, {Barstow}, {Bartolom{\'e}}, {Bassilana},
  {Bauchet}, {Baudesson-Stella}, {Becciani}, {Bellazzini}, {Bernet}, {Bertone},
  {Bianchi}, {Blanco-Cuaresma}, {Boch}, {Bombrun}, {Bossini}, {Bouquillon},
  {Bragaglia}, {Bramante}, {Breedt}, {Bressan}, {Brouillet}, {Bucciarelli},
  {Burlacu}, {Busonero}, {Butkevich}, {Buzzi}, {Caffau}, {Cancelliere},
  {C{\'a}novas}, {Cantat-Gaudin}, {Carballo}, {Carlucci}, {Carnerero},
  {Carrasco}, {Casamiquela}, {Castellani}, {Castro-Ginard}, {Castro Sampol},
  {Chaoul}, {Charlot}, {Chemin}, {Chiavassa}, {Cioni}, {Comoretto}, {Cooper},
  {Cornez}, {Cowell}, {Crifo}, {Crosta}, {Crowley}, {Dafonte}, {Dapergolas},
  {David}, {David}, {de Laverny}, {De Luise}, {De March}, {De Ridder}, {de
  Souza}, {de Teodoro}, {de Torres}, {del Peloso}, {del Pozo}, {Delbo},
  {Delgado}, {Delgado}, {Delisle}, {Di Matteo}, {Diakite}, {Diener},
  {Distefano}, {Dolding}, {Eappachen}, {Edvardsson}, {Enke}, {Esquej}, {Fabre},
  {Fabrizio}, {Faigler}, {Fedorets}, {Fernique}, {Fienga}, {Figueras},
  {Fouron}, {Fragkoudi}, {Fraile}, {Franke}, {Gai}, {Garabato},
  {Garcia-Gutierrez}, {Garc{\'\i}a-Torres}, {Garofalo}, {Gavras}, {Gerlach},
  {Geyer}, {Giacobbe}, {Gilmore}, {Girona}, {Giuffrida}, {Gomel}, {Gomez},
  {Gonzalez-Santamaria}, {Gonz{\'a}lez-Vidal}, {Granvik},
  {Guti{\'e}rrez-S{\'a}nchez}, {Guy}, {Hauser}, {Haywood}, {Helmi}, {Hidalgo},
  {Hilger}, {H{\l}adczuk}, {Hobbs}, {Holland}, {Huckle}, {Jasniewicz},
  {Jonker}, {Juaristi Campillo}, {Julbe}, {Karbevska}, {Kervella}, {Khanna},
  {Kochoska}, {Kontizas}, {Kordopatis}, {Korn}, {Kostrzewa-Rutkowska},
  {Kruszy{\'n}ska}, {Lambert}, {Lanza}, {Lasne}, {Le Campion}, {Le Fustec},
  {Lebreton}, {Lebzelter}, {Leccia}, {Leclerc}, {Lecoeur-Taibi}, {Liao},
  {Licata}, {Lindstr{\o}m}, {Lister}, {Livanou}, {Lobel}, {Madrero Pardo},
  {Managau}, {Mann}, {Marchant}, {Marconi}, {Marcos Santos}, {Marinoni},
  {Marocco}, {Marshall}, {Martin Polo}, {Mart{\'\i}n-Fleitas}, {Masip},
  {Massari}, {Mastrobuono-Battisti}, {Mazeh}, {McMillan}, {Messina},
  {Michalik}, {Millar}, {Mints}, {Molina}, {Molinaro}, {Moln{\'a}r},
  {Montegriffo}, {Mor}, {Morbidelli}, {Morel}, {Morris}, {Mulone}, {Munoz},
  {Muraveva}, {Murphy}, {Musella}, {Noval}, {Ord{\'e}novic}, {Orr{\`u}},
  {Osinde}, {Pagani}, {Pagano}, {Palaversa}, {Palicio}, {Panahi}, {Pawlak},
  {Pe{\~n}alosa Esteller}, {Penttil{\"a}}, {Piersimoni}, {Pineau}, {Plachy},
  {Plum}, {Poggio}, {Poretti}, {Poujoulet}, {Pr{\v{s}}a}, {Pulone}, {Racero},
  {Ragaini}, {Rainer}, {Raiteri}, {Rambaux}, {Ramos}, {Ramos-Lerate}, {Re
  Fiorentin}, {Regibo}, {Reyl{\'e}}, {Ripepi}, {Riva}, {Rixon}, {Robichon},
  {Robin}, {Roelens}, {Rohrbasser}, {Romero-G{\'o}mez}, {Rowell}, {Royer},
  {Rybicki}, {Sadowski}, {Sagrist{\`a} Sell{\'e}s}, {Sahlmann}, {Salgado},
  {Salguero}, {Samaras}, {Sanchez Gimenez}, {Sanna}, {Santove{\~n}a},
  {Sarasso}, {Schultheis}, {Sciacca}, {Segol}, {Segovia}, {S{\'e}gransan},
  {Semeux}, {Shahaf}, {Siddiqui}, {Siebert}, {Siltala}, {Slezak}, {Smart},
  {Solano}, {Solitro}, {Souami}, {Souchay}, {Spagna}, {Spoto}, {Steele},
  {Steidelm{\"u}ller}, {Stephenson}, {S{\"u}veges}, {Szabados}, {Szegedi-Elek},
  {Taris}, {Tauran}, {Taylor}, {Teixeira}, {Thuillot}, {Tonello}, {Torra},
  {Torra}, {Turon}, {Unger}, {Vaillant}, {van Dillen}, {Vanel}, {Vecchiato},
  {Viala}, {Vicente}, {Voutsinas}, {Weiler}, {Wevers}, {Wyrzykowski}, {Yoldas},
  {Yvard}, {Zhao}, {Zorec}, {Zucker}, {Zurbach}, \&
  {Zwitter}}]{2021A&A...649A...1G}
---. 2021, \aap, 649, A1, \dodoi{10.1051/0004-6361/202039657}

\bibitem[{{Garnavich} {et~al.}(2016){Garnavich}, {Tucker}, {Rest}, {Shaya},
  {Olling}, {Kasen}, \& {Villar}}]{2016ApJ...820...23G}
{Garnavich}, P.~M., {Tucker}, B.~E., {Rest}, A., {et~al.} 2016, \apj, 820, 23,
  \dodoi{10.3847/0004-637X/820/1/23}

\bibitem[{Harris {et~al.}(2020)Harris, Millman, van~der Walt, Gommers,
  Virtanen, Cournapeau, Wieser, Taylor, Berg, Smith, Kern, Picus, Hoyer, van
  Kerkwijk, Brett, Haldane, del R{\'{i}}o, Wiebe, Peterson,
  G{\'{e}}rard-Marchant, Sheppard, Reddy, Weckesser, Abbasi, Gohlke, \&
  Oliphant}]{harris2020array}
Harris, C.~R., Millman, K.~J., van~der Walt, S.~J., {et~al.} 2020, Nature, 585,
  357, \dodoi{10.1038/s41586-020-2649-2}

\bibitem[{{Hedges} {et~al.}(2021{\natexlab{a}}){Hedges}, {Luger},
  {Martinez-Palomera}, {Dotson}, \& {Barentsen}}]{2021AJ....162..107H}
{Hedges}, C., {Luger}, R., {Martinez-Palomera}, J., {Dotson}, J., \&
  {Barentsen}, G. 2021{\natexlab{a}}, \aj, 162, 107,
  \dodoi{10.3847/1538-3881/ac0825}

\bibitem[{{Hedges} \& {Mart{\'\i}nez-Palomera}(2021)}]{2021zndo...4784073H}
{Hedges}, C., \& {Mart{\'\i}nez-Palomera}, J. 2021, {SSDataLab/psfmachine:
  Initial Release with Paper}, v1.0.0,  Zenodo, \dodoi{10.5281/zenodo.4784073}

\bibitem[{{Hedges} {et~al.}(2021{\natexlab{b}}){Hedges}, {Saunders}, \&
  {Mart{\'\i}nez-Palomera}}]{2021RNAAS...5..260H}
{Hedges}, C., {Saunders}, N., \& {Mart{\'\i}nez-Palomera}, J.
  2021{\natexlab{b}}, Research Notes of the American Astronomical Society, 5,
  260, \dodoi{10.3847/2515-5172/ac3765}

\bibitem[{Jenkins {et~al.}(2010)Jenkins, Caldwell, Chandrasekaran, Twicken,
  Bryson, Quintana, Clarke, Li, Allen, Tenenbaum, Wu, Klaus, Middour, Cote,
  McCauliff, Girouard, Gunter, Wohler, Sommers, Hall, Uddin, Wu, Bhavsar,
  Cleve, Pletcher, Dotson, Haas, Gilliland, Koch, \& Borucki}]{Jenkins_2010}
Jenkins, J.~M., Caldwell, D.~A., Chandrasekaran, H., {et~al.} 2010, The
  Astrophysical Journal, 713, L87, \dodoi{10.1088/2041-8205/713/2/l87}

\bibitem[{{Jordi} {et~al.}(2010){Jordi}, {Gebran}, {Carrasco}, {de Bruijne},
  {Voss}, {Fabricius}, {Knude}, {Vallenari}, {Kohley}, \&
  {Mora}}]{2010A&A...523A..48J}
{Jordi}, C., {Gebran}, M., {Carrasco}, J.~M., {et~al.} 2010, \aap, 523, A48,
  \dodoi{10.1051/0004-6361/201015441}

\bibitem[{{Kinemuchi} {et~al.}(2012){Kinemuchi}, {Barclay}, {Fanelli},
  {Pepper}, {Still}, \& {Howell}}]{2012PASP..124..963K}
{Kinemuchi}, K., {Barclay}, T., {Fanelli}, M., {et~al.} 2012, \pasp, 124, 963,
  \dodoi{10.1086/667603}

\bibitem[{{Kipping}(2013)}]{2013MNRAS.435.2152K}
{Kipping}, D.~M. 2013, \mnras, 435, 2152, \dodoi{10.1093/mnras/stt1435}

\bibitem[{{Kirk} {et~al.}(2016){Kirk}, {Conroy}, {Pr{\v{s}}a}, {Abdul-Masih},
  {Kochoska}, {Matijevi{\v{c}}}, {Hambleton}, {Barclay}, {Bloemen}, {Boyajian},
  {Doyle}, {Fulton}, {Hoekstra}, {Jek}, {Kane}, {Kostov}, {Latham}, {Mazeh},
  {Orosz}, {Pepper}, {Quarles}, {Ragozzine}, {Shporer}, {Southworth},
  {Stassun}, {Thompson}, {Welsh}, {Agol}, {Derekas}, {Devor}, {Fischer},
  {Green}, {Gropp}, {Jacobs}, {Johnston}, {LaCourse}, {Saetre}, {Schwengeler},
  {Toczyski}, {Werner}, {Garrett}, {Gore}, {Martinez}, {Spitzer}, {Stevick},
  {Thomadis}, {Vrijmoet}, {Yenawine}, {Batalha}, \&
  {Borucki}}]{2016AJ....151...68K}
{Kirk}, B., {Conroy}, K., {Pr{\v{s}}a}, A., {et~al.} 2016, \aj, 151, 68,
  \dodoi{10.3847/0004-6256/151/3/68}

\bibitem[{{Kov{\'a}cs} {et~al.}(2002){Kov{\'a}cs}, {Zucker}, \&
  {Mazeh}}]{2002A&A...391..369K}
{Kov{\'a}cs}, G., {Zucker}, S., \& {Mazeh}, T. 2002, \aap, 391, 369,
  \dodoi{10.1051/0004-6361:20020802}

\bibitem[{Kumar {et~al.}(2019)Kumar, Carroll, Hartikainen, \&
  Martin}]{Kumar2019}
Kumar, R., Carroll, C., Hartikainen, A., \& Martin, O. 2019, Journal of Open
  Source Software, 4, 1143, \dodoi{10.21105/joss.01143}

\bibitem[{{Li} {et~al.}(2019){Li}, {Wang}, {Vink{\'o}}, {Mo}, {Hosseinzadeh},
  {Sand}, {Zhang}, {Lin}, {PTSS/TNTS}, {Zhang}, {Wang}, {Zhang}, {Chen},
  {Xiang}, {Rui}, {Huang}, {Li}, {Zhang}, {Li}, {Baron}, {Derkacy}, {Zhao},
  {Sai}, {Zhang}, {Wang}, {LCO}, {Howell}, {McCully}, {Arcavi}, {Valenti},
  {Hiramatsu}, {Burke}, {KEGS}, {Rest}, {Garnavich}, {Tucker}, {Narayan},
  {Shaya}, {Margheim}, {Zenteno}, {Villar}, {UCSC}, {Dimitriadis}, {Foley},
  {Pan}, {Coulter}, {Fox}, {Jha}, {Jones}, {Kasen}, {Kilpatrick}, {Piro},
  {Riess}, {Rojas-Bravo}, {ASAS-SN}, {Shappee}, {Holoien}, {Stanek}, {Drout},
  {Auchettl}, {Kochanek}, {Brown}, {Bose}, {Bersier}, {Brimacombe}, {Chen},
  {Dong}, {Holmbo}, {Mu{\~n}oz}, {Mutel}, {Post}, {Prieto}, {Shields},
  {Tallon}, {Thompson}, {Vallely}, {Villanueva}, {Pan-STARRS}, {Smartt},
  {Smith}, {Chambers}, {Flewelling}, {Huber}, {Magnier}, {Waters}, {Schultz},
  {Bulger}, {Lowe}, {Willman}, {Konkoly/Texas}, {S{\'a}rneczky}, {P{\'a}l},
  {Wheeler}, {B{\'o}di}, {Bogn{\'a}r}, {Cs{\'a}k}, {Cseh}, {Cs{\"o}rnyei},
  {Hanyecz}, {Ign{\'a}cz}, {Kalup}, {K{\"o}nyves-T{\'o}th}, {Kriskovics},
  {Ordasi}, {Rajmon}, {S{\'o}dor}, {Szab{\'o}}, {Szak{\'a}ts}, {Zsidi},
  {Arizona}, {Milne}, {Andrews}, {Smith}, {Bilinski}, {Swift}, {Brown},
  {ePESSTO}, {Nordin}, {Williams}, {Galbany}, {Palmerio}, {Hook}, {Inserra},
  {Maguire}, {Cartier}, {Razza}, {Guti{\'e}rrez}, {North Carolina}, {Hermes},
  {Reding}, {Kaiser}, {ATLAS}, {Tonry}, {Heinze}, {Denneau}, {Weiland},
  {Stalder}, {K2 Mission Team}, {Barentsen}, {Dotson}, {Barclay},
  {Gully-Santiago}, {Hedges}, {Cody}, {Howell}, {Kepler Spacecraft Team},
  {Coughlin}, {Van Cleve}, {Cardoso}, {Larson}, {McCalmont-Everton},
  {Peterson}, {Ross}, {Reedy}, {Osborne}, {McGinn}, {Kohnert}, {Migliorini},
  {Wheaton}, {Spencer}, {Labonde}, {Castillo}, {Beerman}, {Steward}, {Hanley},
  {Larsen}, {Gangopadhyay}, {Kloetzel}, {Weschler}, {Nystrom}, {Moffatt},
  {Redick}, {Griest}, {Packard}, {Muszynski}, {Kampmeier}, {Bjella}, {Flynn},
  \& {Elsaesser}}]{2019ApJ...870...12L}
{Li}, W., {Wang}, X., {Vink{\'o}}, J., {et~al.} 2019, \apj, 870, 12,
  \dodoi{10.3847/1538-4357/aaec74}

\bibitem[{{Lightkurve Collaboration} {et~al.}(2018){Lightkurve Collaboration},
  {Cardoso}, {Hedges}, {Gully-Santiago}, {Saunders}, {Cody}, {Barclay}, {Hall},
  {Sagear}, {Turtelboom}, {Zhang}, {Tzanidakis}, {Mighell}, {Coughlin}, {Bell},
  {Berta-Thompson}, {Williams}, {Dotson}, \& {Barentsen}}]{2018ascl.soft12013L}
{Lightkurve Collaboration}, {Cardoso}, J.~V.~d.~M., {Hedges}, C., {et~al.}
  2018, {Lightkurve: Kepler and TESS time series analysis in Python},
  Astrophysics Source Code Library.
\newblock \doeprint{1812.013}

\bibitem[{{Lindegren} {et~al.}(2018){Lindegren}, {Hern{\'a}ndez}, {Bombrun},
  {Klioner}, {Bastian}, {Ramos-Lerate}, {de Torres}, {Steidelm{\"u}ller},
  {Stephenson}, {Hobbs}, {Lammers}, {Biermann}, {Geyer}, {Hilger}, {Michalik},
  {Stampa}, {McMillan}, {Casta{\~n}eda}, {Clotet}, {Comoretto}, {Davidson},
  {Fabricius}, {Gracia}, {Hambly}, {Hutton}, {Mora}, {Portell}, {van Leeuwen},
  {Abbas}, {Abreu}, {Altmann}, {Andrei}, {Anglada}, {Balaguer-N{\'u}{\~n}ez},
  {Barache}, {Becciani}, {Bertone}, {Bianchi}, {Bouquillon}, {Bourda},
  {Br{\"u}semeister}, {Bucciarelli}, {Busonero}, {Buzzi}, {Cancelliere},
  {Carlucci}, {Charlot}, {Cheek}, {Crosta}, {Crowley}, {de Bruijne}, {de
  Felice}, {Drimmel}, {Esquej}, {Fienga}, {Fraile}, {Gai}, {Garralda},
  {Gonz{\'a}lez-Vidal}, {Guerra}, {Hauser}, {Hofmann}, {Holl}, {Jordan},
  {Lattanzi}, {Lenhardt}, {Liao}, {Licata}, {Lister}, {L{\"o}ffler},
  {Marchant}, {Martin-Fleitas}, {Messineo}, {Mignard}, {Morbidelli}, {Poggio},
  {Riva}, {Rowell}, {Salguero}, {Sarasso}, {Sciacca}, {Siddiqui}, {Smart},
  {Spagna}, {Steele}, {Taris}, {Torra}, {van Elteren}, {van Reeven}, \&
  {Vecchiato}}]{2018A&A...616A...2L}
{Lindegren}, L., {Hern{\'a}ndez}, J., {Bombrun}, A., {et~al.} 2018, \aap, 616,
  A2, \dodoi{10.1051/0004-6361/201832727}

\bibitem[{{Luger} {et~al.}(2019){Luger}, {Agol}, {Foreman-Mackey}, {Fleming},
  {Lustig-Yaeger}, \& {Deitrick}}]{2019AJ....157...64L}
{Luger}, R., {Agol}, E., {Foreman-Mackey}, D., {et~al.} 2019, \aj, 157, 64,
  \dodoi{10.3847/1538-3881/aae8e5}

\bibitem[{Martínez-Palomera(2021)}]{jorge_martinez_palomera_2021_5062871}
Martínez-Palomera, J. 2021, {jorgemarpa/kepler-apertures: Kepler-apertures
  first release}, v0.1.0,  Zenodo, \dodoi{10.5281/zenodo.5062871}

\bibitem[{{McClure} {et~al.}(2021){McClure}, {Soares}, {Mathieu}, \&
  {Moore}}]{2021AAS...23714006M}
{McClure}, R.~L., {Soares}, M., {Mathieu}, R., \& {Moore}, C. 2021, in American
  Astronomical Society Meeting Abstracts, Vol.~53, American Astronomical
  Society Meeting Abstracts, 140.06

\bibitem[{Montet {et~al.}(2017)Montet, Tovar, \& Foreman-Mackey}]{Montet_2017}
Montet, B.~T., Tovar, G., \& Foreman-Mackey, D. 2017, The Astrophysical
  Journal, 851, 116, \dodoi{10.3847/1538-4357/aa9e00}

\bibitem[{{Olling} {et~al.}(2015){Olling}, {Mushotzky}, {Shaya}, {Rest},
  {Garnavich}, {Tucker}, {Kasen}, {Margheim}, \&
  {Filippenko}}]{2015Natur.521..332O}
{Olling}, R.~P., {Mushotzky}, R., {Shaya}, E.~J., {et~al.} 2015, \nat, 521,
  332, \dodoi{10.1038/nature14455}

\bibitem[{{Paxton} {et~al.}(2011){Paxton}, {Bildsten}, {Dotter}, {Herwig},
  {Lesaffre}, \& {Timmes}}]{2011ApJS..192....3P}
{Paxton}, B., {Bildsten}, L., {Dotter}, A., {et~al.} 2011, \apjs, 192, 3,
  \dodoi{10.1088/0067-0049/192/1/3}

\bibitem[{{Paxton} {et~al.}(2013){Paxton}, {Cantiello}, {Arras}, {Bildsten},
  {Brown}, {Dotter}, {Mankovich}, {Montgomery}, {Stello}, {Timmes}, \&
  {Townsend}}]{2013ApJS..208....4P}
{Paxton}, B., {Cantiello}, M., {Arras}, P., {et~al.} 2013, \apjs, 208, 4,
  \dodoi{10.1088/0067-0049/208/1/4}

\bibitem[{{Paxton} {et~al.}(2015){Paxton}, {Marchant}, {Schwab}, {Bauer},
  {Bildsten}, {Cantiello}, {Dessart}, {Farmer}, {Hu}, {Langer}, {Townsend},
  {Townsley}, \& {Timmes}}]{2015ApJS..220...15P}
{Paxton}, B., {Marchant}, P., {Schwab}, J., {et~al.} 2015, \apjs, 220, 15,
  \dodoi{10.1088/0067-0049/220/1/15}

\bibitem[{Salvatier {et~al.}(2016)Salvatier, Wiecki, \&
  Fonnesbeck}]{10.7717/peerj-cs.55}
Salvatier, J., Wiecki, T.~V., \& Fonnesbeck, C. 2016, PeerJ Computer Science,
  2, e55, \dodoi{10.7717/peerj-cs.55}

\bibitem[{Sandford \& Kipping(2017)}]{Sandford_2017}
Sandford, E., \& Kipping, D. 2017, The Astronomical Journal, 154, 228,
  \dodoi{10.3847/1538-3881/aa94bf}

\bibitem[{{Schlafly} \& {Finkbeiner}(2011)}]{2011ApJ...737..103S}
{Schlafly}, E.~F., \& {Finkbeiner}, D.~P. 2011, \apj, 737, 103,
  \dodoi{10.1088/0004-637X/737/2/103}

\bibitem[{{Schlegel} {et~al.}(1998){Schlegel}, {Finkbeiner}, \&
  {Davis}}]{1998ApJ...500..525S}
{Schlegel}, D.~J., {Finkbeiner}, D.~P., \& {Davis}, M. 1998, \apj, 500, 525,
  \dodoi{10.1086/305772}

\bibitem[{{Smith} {et~al.}(2012){Smith}, {Stumpe}, {Van Cleve}, {Jenkins},
  {Barclay}, {Fanelli}, {Girouard}, {Kolodziejczak}, {McCauliff}, {Morris}, \&
  {Twicken}}]{2012PASP..124.1000S}
{Smith}, J.~C., {Stumpe}, M.~C., {Van Cleve}, J.~E., {et~al.} 2012, \pasp, 124,
  1000, \dodoi{10.1086/667697}

\bibitem[{{Stello} {et~al.}(2010){Stello}, {Basu}, {Bruntt}, {Mosser},
  {Stevens}, {Brown}, {Christensen-Dalsgaard}, {Gilliland}, {Kjeldsen},
  {Arentoft}, {Ballot}, {Barban}, {Bedding}, {Chaplin}, {Elsworth},
  {Garc{\'\i}a}, {Goupil}, {Hekker}, {Huber}, {Mathur}, {Meibom},
  {Sangaralingam}, {Baldner}, {Belkacem}, {Biazzo}, {Brogaard}, {Su{\'a}rez},
  {D'Antona}, {Demarque}, {Esch}, {Gai}, {Grundahl}, {Lebreton}, {Jiang},
  {Jevtic}, {Karoff}, {Miglio}, {Molenda-{\.Z}akowicz}, {Montalb{\'a}n},
  {Noels}, {Roca Cort{\'e}s}, {Roxburgh}, {Serenelli}, {Silva Aguirre},
  {Sterken}, {Stine}, {Szab{\'o}}, {Weiss}, {Borucki}, {Koch}, \&
  {Jenkins}}]{2010ApJ...713L.182S}
{Stello}, D., {Basu}, S., {Bruntt}, H., {et~al.} 2010, \apjl, 713, L182,
  \dodoi{10.1088/2041-8205/713/2/L182}

\bibitem[{{Stello} {et~al.}(2011){Stello}, {Huber}, {Kallinger}, {Basu},
  {Mosser}, {Hekker}, {Mathur}, {Garc{\'\i}a}, {Bedding}, {Kjeldsen},
  {Gilliland}, {Verner}, {Chaplin}, {Benomar}, {Meibom}, {Grundahl},
  {Elsworth}, {Molenda-{\.Z}akowicz}, {Szab{\'o}}, {Christensen-Dalsgaard},
  {Tenenbaum}, {Twicken}, \& {Uddin}}]{2011ApJ...737L..10S}
{Stello}, D., {Huber}, D., {Kallinger}, T., {et~al.} 2011, \apjl, 737, L10,
  \dodoi{10.1088/2041-8205/737/1/L10}

\bibitem[{{The Theano Development Team} {et~al.}(2016){The Theano Development
  Team}, {Al-Rfou}, {Alain}, {Almahairi}, {Angermueller}, {Bahdanau}, {Ballas},
  {Bastien}, {Bayer}, {Belikov}, {Belopolsky}, {Bengio}, {Bergeron},
  {Bergstra}, {Bisson}, {Bleecher Snyder}, {Bouchard}, {Boulanger-Lewandowski},
  {Bouthillier}, {de Br{\'e}bisson}, {Breuleux}, {Carrier}, {Cho}, {Chorowski},
  {Christiano}, {Cooijmans}, {C{\^o}t{\'e}}, {C{\^o}t{\'e}}, {Courville},
  {Dauphin}, {Delalleau}, {Demouth}, {Desjardins}, {Dieleman}, {Dinh},
  {Ducoffe}, {Dumoulin}, {Ebrahimi Kahou}, {Erhan}, {Fan}, {Firat}, {Germain},
  {Glorot}, {Goodfellow}, {Graham}, {Gulcehre}, {Hamel}, {Harlouchet}, {Heng},
  {Hidasi}, {Honari}, {Jain}, {Jean}, {Jia}, {Korobov}, {Kulkarni}, {Lamb},
  {Lamblin}, {Larsen}, {Laurent}, {Lee}, {Lefrancois}, {Lemieux},
  {L{\'e}onard}, {Lin}, {Livezey}, {Lorenz}, {Lowin}, {Ma}, {Manzagol},
  {Mastropietro}, {McGibbon}, {Memisevic}, {van Merri{\"e}nboer}, {Michalski},
  {Mirza}, {Orlandi}, {Pal}, {Pascanu}, {Pezeshki}, {Raffel}, {Renshaw},
  {Rocklin}, {Romero}, {Roth}, {Sadowski}, {Salvatier}, {Savard},
  {Schl{\"u}ter}, {Schulman}, {Schwartz}, {Vlad Serban}, {Serdyuk},
  {Shabanian}, {Simon}, {Spieckermann}, {Ramana Subramanyam}, {Sygnowski},
  {Tanguay}, {van Tulder}, {Turian}, {Urban}, {Vincent}, {Visin}, {de Vries},
  {Warde-Farley}, {Webb}, {Willson}, {Xu}, {Xue}, {Yao}, {Zhang}, \&
  {Zhang}}]{2016arXiv160502688T}
{The Theano Development Team}, {Al-Rfou}, R., {Alain}, G., {et~al.} 2016, arXiv
  e-prints, arXiv:1605.02688.
\newblock \doarXiv{1605.02688}

\bibitem[{{Thompson} {et~al.}(2016){Thompson}, {Fraquelli}, {Van Cleve}, \&
  {Caldwell}}]{2016ksci.rept....9T}
{Thompson}, S.~E., {Fraquelli}, D., {Van Cleve}, J.~E., \& {Caldwell}, D.~A.
  2016, {Kepler Archive Manual}, Kepler Science Document KDMC-10008-006

\bibitem[{Thompson {et~al.}(2018)Thompson, Coughlin, Hoffman, Mullally,
  Christiansen, Burke, Bryson, Batalha, Haas, Catanzarite, Rowe, Barentsen,
  Caldwell, Clarke, Jenkins, Li, Latham, Lissauer, Mathur, Morris, Seader,
  Smith, Klaus, Twicken, Cleve, Wohler, Akeson, Ciardi, Cochran, Henze, Howell,
  Huber, Pr{\v{s}}a, Ram{\'{\i}}rez, Morton, Barclay, Campbell, Chaplin,
  Charbonneau, Christensen-Dalsgaard, Dotson, Doyle, Dunham, Dupree, Ford,
  Geary, Girouard, Isaacson, Kjeldsen, Quintana, Ragozzine, Shabram, Shporer,
  Aguirre, Steffen, Still, Tenenbaum, Welsh, Wolfgang, Zamudio, Koch, \&
  Borucki}]{Thompson_2018}
Thompson, S.~E., Coughlin, J.~L., Hoffman, K., {et~al.} 2018, The Astrophysical
  Journal Supplement Series, 235, 38, \dodoi{10.3847/1538-4365/aab4f9}

\bibitem[{{Van Cleve} \& {Caldwell}(2016)}]{2016ksci.rept....1V}
{Van Cleve}, J.~E., \& {Caldwell}, D.~A. 2016, {Kepler Instrument Handbook},
  Kepler Science Document KSCI-19033-002

\bibitem[{{Van Cleve} {et~al.}(2016){Van Cleve}, {Christiansen}, {Jenkins},
  {Caldwell}, {Barclay}, {Bryson}, {Burke}, {Cambell}, {Catanzarite}, {Clarke},
  {Coughlin}, {Girouard}, {Haas}, {Klaus}, {Kolodziejczak}, {Li}, {McCauliff},
  {Morris}, {Mullally}, {Quintana}, {Rowe}, {Sabale}, {Seader}, {Smith},
  {Still}, {Tenenbaum}, {Thompson}, {Twicken}, {Kamal Uddin}, \&
  {Zamudio}}]{2016ksci.rept....2V}
{Van Cleve}, J.~E., {Christiansen}, J.~L., {Jenkins}, J.~M., {et~al.} 2016,
  {Kepler Data Characteristics Handbook}, Kepler Science Document
  KSCI-19040-005

\bibitem[{{Vanderburg} \& {Johnson}(2014)}]{2014PASP..126..948V}
{Vanderburg}, A., \& {Johnson}, J.~A. 2014, \pasp, 126, 948,
  \dodoi{10.1086/678764}

\bibitem[{Virtanen {et~al.}(2020)Virtanen, Gommers, Oliphant, Haberland, Reddy,
  Cournapeau, Burovski, Peterson, Weckesser, Bright, {van der Walt}, Brett,
  Wilson, Millman, Mayorov, Nelson, Jones, Kern, Larson, Carey, Polat, Feng,
  Moore, {VanderPlas}, Laxalde, Perktold, Cimrman, Henriksen, Quintero, Harris,
  Archibald, Ribeiro, Pedregosa, {van Mulbregt}, \& {SciPy 1.0
  Contributors}}]{2020SciPy-NMeth}
Virtanen, P., Gommers, R., Oliphant, T.~E., {et~al.} 2020, Nature Methods, 17,
  261, \dodoi{10.1038/s41592-019-0686-2}

\bibitem[{{Wu} {et~al.}(2010){Wu}, {Twicken}, {Tenenbaum}, {Clarke}, {Li},
  {Quintana}, {Allen}, {Chandrasekaran}, {Jenkins}, {Caldwell}, {Wohler},
  {Girouard}, {McCauliff}, {Cote}, \& {Klaus}}]{2010SPIE.7740E..19W}
{Wu}, H., {Twicken}, J.~D., {Tenenbaum}, P., {et~al.} 2010, in Society of
  Photo-Optical Instrumentation Engineers (SPIE) Conference Series, Vol. 7740,
  Software and Cyberinfrastructure for Astronomy, ed. N.~M. {Radziwill} \&
  A.~{Bridger}, 774019, \dodoi{10.1117/12.856630}

\end{thebibliography}



\end{document}